\documentclass[journal]{IEEEtran}  

\usepackage{arydshln}
\usepackage{setspace}
\usepackage{multirow}
\usepackage{booktabs}

\IEEEoverridecommandlockouts
\usepackage{graphics} 
\usepackage{epsfig} 
\usepackage{times} 

\usepackage{algorithm}
\usepackage[noend]{algpseudocode}

\makeatletter
\def\BState{\State\hskip-\ALG@thistlm}
\makeatother

\usepackage{amssymb}

\usepackage{amsthm}
\usepackage{amsmath} 
\usepackage{amssymb}  
\usepackage{epstopdf}
\usepackage{epsfig}
\usepackage{subfigure}
\newcommand{\m}{\boldsymbol}
\usepackage{tikz}

\makeatother
\newtheorem{mydef}{Definition}

\newtheorem{rem}{Remark}

\newtheorem{assumption}{Assumption}
\usepackage{amsmath}
\usepackage{amssymb}
\usepackage{textcase}
\usepackage{soul}
\usepackage{indentfirst}

\DeclareMathOperator{\rank}{rank}

\DeclareMathOperator*{\maximize}{maximize}
\DeclareMathOperator{\subjectto}{subject\ to}
\DeclareMathOperator*{\sat}{sat}
\newcommand{\mat}[1]{\boldsymbol{#1}}

\DeclareMathOperator*{\sqq}{square}
\DeclareMathOperator*{\sawtooth}{sawtooth}

\newcommand{\boxalign}[2][0.44\textwidth]{
	\par\noindent\tikzstyle{mybox} = [draw=black,inner sep=10pt]
	\begin{center}\begin{tikzpicture}
		\node [mybox] (box){%
			\begin{minipage}{#1}{\vspace{-5mm}#2}\end{minipage}
		};
		\end{tikzpicture}\end{center}
}

\newcommand{\normof}[1]{\|#1\|}

\newcommand{\bmat}[1]{\begin{bmatrix} #1 \end{bmatrix}}

\providecommand{\mA}{\ensuremath{\mat{A}}}
\providecommand{\mB}{\ensuremath{\mat{B}}}
\providecommand{\mC}{\ensuremath{\mat{C}}}

\providecommand{\mE}{\ensuremath{\mat{E}}}
\providecommand{\mF}{\ensuremath{\mat{F}}}

\providecommand{\mH}{\ensuremath{\mat{H}}}
\providecommand{\mI}{\ensuremath{\mat{I}}}

\providecommand{\mL}{\ensuremath{\mat{L}}}

\providecommand{\mP}{\ensuremath{\mat{P}}}
\providecommand{\mQ}{\ensuremath{\mat{Q}}}

\providecommand{\mU}{\ensuremath{\mat{U}}}
\providecommand{\mV}{\ensuremath{\mat{V}}}
\providecommand{\mW}{\ensuremath{\mat{W}}}

\providecommand{\mY}{\ensuremath{\mat{Y}}}

\usepackage[noadjust]{cite}

\begin{document}

\title{Risk Mitigation for Dynamic State Estimation Against Cyber Attacks and Unknown Inputs}

\author{Ahmad F. Taha,~\IEEEmembership{Member,~IEEE, }
	Junjian~Qi,~\IEEEmembership{Member,~IEEE, }
	Jianhui Wang,~\IEEEmembership{Senior Member,~IEEE }
	\\ and Jitesh H. Panchal,~\IEEEmembership{Member,~IEEE}
	\thanks{This work was supported by the U.S. Department of Energy Office of Electricity Delivery and Energy Reliability and by the Smart Grid Security
		Research Grant from the UTSA Office of the Vice President of Research. Paper no. TSG-00921-2015.
		
       A. F. Taha is with the Department of Electrical and Computer Engineering, the University of Texas at San Antonio, San Antonio, TX 78249 (e-mail: ahmad.taha@utsa.edu).
	   
	   J. Qi and J. Wang are with the Energy Systems Division, Argonne National Laboratory, Argonne, IL 60439 USA (e-mails: jqi@anl.gov; jianhui.wang@anl.gov).
	   
	   J. H. Panchal is with the School of Mechanical Engineering, Purdue University, West Lafayette, IN 47907 USA (e-mail: panchal@purdue.edu).}
}

\maketitle

\begin{abstract}
	Phasor measurement units (PMUs) can be effectively utilized for the monitoring and control of the power grid. As the cyber-world becomes increasingly embedded into power grids, the risks of this inevitable evolution become serious. In this paper, we present a risk mitigation strategy, based on dynamic state estimation, to eliminate threat levels from the grid's unknown inputs and potential cyber-attacks. The strategy requires (a) the potentially incomplete knowledge of power system models and parameters and (b) real-time PMU measurements. 
	First, we utilize a dynamic state estimator for higher order depictions of power system dynamics for simultaneous state and unknown inputs estimation. Second, estimates of cyber-attacks are obtained through an attack detection algorithm. Third, the estimation and detection components are seamlessly utilized in an optimization framework to determine the most impacted PMU measurements. Finally, a risk mitigation strategy is proposed to guarantee the elimination of threats from attacks, ensuring the observability of the power system through available, \textit{safe} measurements. Case studies are included to validate the proposed approach. Insightful suggestions, extensions, and open problems are also posed.
\end{abstract}


\section*{Acronyms}
\addcontentsline{toc}{section}{Acronyms}
\begin{IEEEdescription}[\IEEEusemathlabelsep\IEEEsetlabelwidth{$V_1,V_2,V_3,V_4$}]
	\item[CA] Cyber-attack.
	\item[DRMA] Dynamic risk mitigation algorithm.
	\item[DRMOP] Dynamic risk mitigation opt. problem.
	\item[DSE] Dynamic state estimation.
	\item[ILP] Integer linear program.
	\item[LMI] Linear matrix inequality.
	\item[PMU] Phasor measurement unit.
	\item[SMO] Sliding mode observer.
	\item[UI] Unknown input.	
	\item[WDTL] Weighted deterministic threat level.
\end{IEEEdescription}
\section*{Nomenclature}
\addcontentsline{toc}{section}{Nomenclature}
\begin{IEEEdescription}[\IEEEusemathlabelsep\IEEEsetlabelwidth{$V_1,V_2,V_3,V_4$}]
	\item[$\m x,\hat{\m x}$] States and the estimate.
	\item[$\m e$] State estimation error, i.e., $\m x -\hat{\m x}$.
	\item[$\m y_q,\hat{\m y}_q$] Measurements and the estimate.		\item[$\m w,\hat{\m w}$] Unknown input vector and its estimate.	
				\item[$\m v_q,\hat{\m v}_q$] CA vector and its estimate.	
			\item[$\m l$] Column vector of the attack detection filter.	
			\item[$\m r$] Residual of the attack detection filter.
		\item[$\m z$] WDTL vector.
		\item[$\m \pi$] Vector of binary decision variables which is equal to $1$ if the $i$th PMU measurement is used for state estimation and $0$ otherwise.
			\item[$\m \lambda$] Vector of eigenvalues of $\mA$.
		\item[$\mA$] Linearized system state matrix.
	\item[$\mB_w$] Weight distribution matrix for UIs.
	\item[$\mC_q$] Linearized power system output matrix.
	\item [$\mathcal{O}$] Observability matrix.
	\item[$\mL_q,\mF_q,\mP$]  Sliding mode observer design matrices.
	\item [$\boldsymbol{\overline{Y}},\boldsymbol{\overline{Y}}_i$] Admittance matrix of the reduced network consisting of generators and its $i$th row. 
    \item [$\m \Psi_R, \m \Psi_I$] Column vector of all generators' real and imaginary part of the voltage source on system reference frame. 
	\item[$\alpha_i$] Cost weight for activating or deactivating the $i$th PMU measurement.
	\item[$\beta_i$] Weight of the $i$th PMU measurement.
	\item [$\gamma_i$] Residual threshold of the $i$th measurement.
	\item[$\eta,\nu$] SMO gain and smoothing constants.
	\item[$\zeta$] Rank of $\mB_w$.
	\item [$\delta$] Rotor angle in rad. 
	\item [$\omega, \omega_0,\omega_f$] Rotor speed, rated rotor speed, and rotor speed set point in rad/s.
	\item [$\omega_e$] Rotor speed deviation in pu.
	\item [$E_{fd},E_{fd}^0$] Internal field voltage and its initial value in pu. 
	\item [$E_t$] Terminal voltage phasor.
	\item [$E_T^0$] Initial machine terminal voltage.
	\item [$e_q, e_d$] Terminal voltage at $q$ axis and $d$ axis in pu. 
	\item [$e'_q, e'_d$] Transient voltage at $q$ axis and $d$ axis in pu. 
	\item [$e_R, e_I$] Real and imaginary part of the terminal voltage phasor. 
	\item [$exc^{1,2,3}$] Internally set exciter constants.
	\item [$\mathcal{G}_P$] Set of generators where PMUs are installed. 
	\item [$H$] Generator inertia constant in second. 
	\item [$I_t$] Terminal current phasor. 
	\item [$i_q, i_d$] Current at $q$ and $d$ axes in pu.
	\item [$i_R, i_I$] Real and imaginary part of the terminal current phasor in pu. 
	\item [$K_A$] Voltage regulator gain.
	\item [$K_D$] Damping factor in pu. 
	\item [$K_E$] Exciter constant.
	\item [$K_F$] Stabilizer gain.
	\item [$P_e$] Electric power in pu. 
	\item [$P_{m}^0$] Initial mechanical input power.
	\item[$q,n_g,n_w$] Number of PMUs, generators, and UIs.
	\item [$R_f$] Stabilizing transformer state variable.
	\item [$S_B, S_N$] System base and generator base MVA. 
	\item [${tg}_1, {tg}_2, {tg}_3$] Governor, servo, and reheater variables.
	\item [$T_A, T_e, T_F$] Voltage regulator, exciter, and stabilizer time constants.
	\item [$T_m, T_e$] Mechanical torque and electric air-gap torque in pu.
	\item [$T^{\max}$] Maximum power order.
	\item [$T'_{q0}, T'_{d0}$] Open-circuit time constants for $q$ and $d$ axes in seconds.
	\item [$T_s, T_c$] Servo and HP turbine time constants.
	\item [$T_3, T_4, T_5$] Transient gain time constant, time constant to set HP ratio, and reheater time constant.
	\item [$V_A, V_R$] Regulator output voltage in pu.
	\item [$V_{FB}$] Feedback from stabilizing transformer.
	\item [$V_{TR}$] Voltage transducer output in pu.
	\item [$x_{q}, x_{d}$] Synchronous reactance at $q,d$ axes in pu.
	\item [$x'_{q}, x'_{d}$] Transient reactance at $q,d$ axes in pu.  
	\item [$1/r$] Steady state gain.
	\item [$\textrm{sgn}(\cdot)$] Signum function.
\end{IEEEdescription}


\section{Introduction and Motivation}\label{sec:intro}

\IEEEPARstart{C}{lassical} supervisory control and data acquisition (SCADA) systems have become insufficient to guarantee real-time protection of power systems' assets. Consequently, the research and development of wide area measurement systems (WAMS) have significantly increased. By utilizing the phasor measurement units (PMUs), WAMS technologies enable near real-time monitoring of the system, hence empowering a more accurate depiction of the power-grid's cyber-physical status---and improved grid control. 

Recently, the National Electric Sector Cybersecurity Organization Resource (NESCOR) investigated many cybersecurity failure scenarios, which are defined as ``realistic event in which the failure to maintain confidentiality, integrity, and/or availability of sector cyber assets creates a negative impact on the generation, transmission, and/or delivery of power"~\cite{NESCOR2014}. Among these failure scenarios the following two wide-area monitoring, protection, and control (WAMPAC) scenarios motivate the research in this paper:
\begin{itemize}
	\item WAMPAC.4: Measurement Data Compromised due to PDC\footnote{A single PMU transmits measurements to a phasor data concentrator (PDC), and then to a super PDC, through a wireless communication network based on the NASPInet architecture~\cite{Wang2013}.} Authentication Compromise;
	\item WAMPAC.6: Communications Compromised between PMUs and Control Center.
\end{itemize}
Specifically, we consider the problem of attacking PMU measurements by compromising the signals sent to the control center. The two aforementioned scenarios are related in the sense that compromising the communication between PMUs, PDCs, and control center can include alteration of PMU data. In addition to that, we assume that the dynamics of a power system are inherently uncertain. 

\section{Literature Review \& Gaps, Paper Objectives}

The most widely studied static state estimation (SSE) \cite{se1,se2,se3,se5,qi2012power} cannot capture the dynamics of power systems well due to its dependency on slow update rates of SCADA systems. In contrast, dynamic state estimation (DSE) enabled by PMUs can provide accurate dynamic states of the system, and will predictably play a critical role in achieving real-time wide-area monitoring, protection, and control. DSE has been implemented by extended Kalman filter \cite{ekf1,ghahremani2011dynamic}, unscented Kalman filter \cite{ghahremani2011online,pwukf1,singh2014decentralized,sun2016power}, square-root unscented Kalman filter \cite{qi2015dynamic}, extended particle filter \cite{zhou}, and ensemble Kalman filter \cite{zhou2015dynamic}. 
Other dynamical state observers for power systems with unknown inputs (UI) or under cyber-attacks (CA) have also been developed, as in \cite{KalsiDisser,Teix2010,qi2016comparing}. 

DSE requires a reliable dynamic model of the power system, which can be based on post-validation of the dynamic model and calibration the parameters of generators, as in \cite{huang2013generator,hajnoroozi2015generating,ariff2015estimating}; however, there is still a gap between the model and actual power system physics. Assuming that the dynamical models are perfectly accurate can generate sub-optimal estimation laws. In this paper we discuss how this discrepancy can be systematically addressed by the estimation of UIs.

Detecting and isolating CAs in cyber-physical systems generally, and smart-grids specifically, has received immense attention.
Liu \textit{et al.} present a new class of attacks, called false data injection attacks, targeted against SSE in power networks~\cite{Liu2009},  
and show that an attacker can launch successful attacks to alter state estimate. 
In~\cite{Pasq2011,Pasq2013}, the authors propose a generic framework for attack detection, metrics on controllability and observability, and centralized \& distributed attack detection monitors, for a linear time-invariant representation of power systems. 
The reader is referred to~\cite{Pasq2015} for a survey on different types of CAs and attack detection and identification methods that are mainly based on control-theoretic foundations and to~\cite{Wang2013} for a survey on cyber-security in smart grids.

In \cite{scpse}, a security-oriented cyber-physical state estimation system is proposed to identify the maliciously modified set of measurements 
by a combinatorial-based bad-data detection method based on power measurements and cyber security state estimation result. However, this work is still on SSE which is significantly different from the DSE discussed in this paper. 

In~\cite{Amir2015}, Mousavian \textit{et al.} present a probabilistic risk mitigation model for CAs against PMU networks, in which a mixed integer linear programming~(MILP) is formulated that incorporates the derived threat levels of PMUs into a risk-mitigation technique. 
In this MILP, the binary variables determine whether a certain PMU shall be kept connected to the PMU network or removed, while minimizing the maximum threat level for all connected PMUs~\cite{Amir2015}.
However, the estimation problem with PMUs is not considered---there is no connection between the real-time states of the power system and the threat levels.
In this paper, we evaluate the measured and estimated PMU signals, as well as the estimates of UIs and CAs, as an essential deterministic component in the decision-making problem that determines which PMU measurements should be disconnected from the estimation process. 

The objective of this paper is to develop a framework that (a) leverages PMU data to detect disturbances and attacks against a power network and (b) enables secure estimation of power system states, UIs, and CAs. 
In Section~\ref{sec:SystemModel}, we present the power system model used for DSE. 
The physical meaning of the UIs and CAs is discussed in Section~\ref{sec:UIAttackModel}. 
The dynamical models of state-observer is discussed in Section~\ref{sec:StateEstimation}. 
Given a dynamical observer, closed-form estimates for vectors of UIs and CAs, as well as an attack detection filter are all discussed in Section~\ref{sec:AttackEst}. Utilizing the aforementioned estimates, a dynamic risk mitigation algorithm is formulated in Section~\ref{sec:AttackMitig}. 
Section~\ref{sec:flowchart} summarizes the overall solution scheme.
In Section~\ref{sec:sim}, numerical results on the 16-machine 68-bus power system are presented to validate the proposed risk mitigation approach. Finally closing remarks and open research problems are discussed in Section~\ref{sec:concl}.

\allowdisplaybreaks
\section{Power System Model}~\label{sec:SystemModel}
Here, we review the nonlinear dynamics and small-signal linearized representation of a power system.
\subsection{Nonlinear Dynamics of the Power System}~\label{sec:NLDyn}
The fast sub-transient dynamics and saturation effects are ignored and each of the $n_g$ generators is described by the two-axis transient model with an IEEE Type DC1 excitation system and a simplified turbine-governor system \cite{sauer1998power}:
\begin{eqnarray}~\label{gen model}
\hspace{-0.3cm}\left\{
\arraycolsep=1.4pt\def\arraystretch{1.8}
\begin{array}{ll}
\dot{\delta_i}=\omega_i-\omega_0 \\
\dot{\omega}_i=\dfrac{\omega_0}{2H_i}\left( T_{m_{i}}-T_{e_{i}}-\dfrac{K_{D_{i}}}{\omega_0}\left(\omega_i-\omega_0\right)\right) \\
\dot{e}'_{q_{i}}=\dfrac{1}{T'_{d_{0_{i}}}}\left(E_{fd_{i}}-e'_{q_{i}}-(x_{d_{i}}-x'_{d_{i}})i_{d_{i}}\right) \\
\dot{e}'_{d_{i}}=\dfrac{1}{T'_{q_{0_{i}}}}\left(-e'_{di}+(x_{q_{i}} -x'_{q_{i}} )i_{q_{i}}\right) \\
\dot{V}_{R_{i}}=\dfrac{1}{T_{A_{i}}}(-V_{R_{i}}+K_{A_{i}} V_{A_{i}}) \\
\dot{E}_{f_{d_{i}}}=\dfrac{1}{T_{e_{i}}}(V_{R_{i}}-K_{E_{i}}E_{fd_{i}}-S_{E_{i}}) \\
\dot{R}_{f_{i}}=\dfrac{1}{T_{F_{i}}}(-R_{f_{i}}+E_{fd_{i}}) \\
\dot{tg}_{1_{i}}=\dfrac{1}{T_{si}}(D_i-tg_{1_{i}}) \\
\dot{tg}_{2_{i}}=\dfrac{1}{T_{c_{i}}}\left(\left(1-\dfrac{T_{3_{i}}}{T_{c_{i}}}\right)\,{tg}_{1_{i}}-{tg}_{2_{i}}\right) \\
\dot{tg}_{3_{i}}=\dfrac{1}{T_{5_{i}}}\left(\left(\dfrac{T_{3_{i}}}{T_{c_{i}}}{tg}_{1_{i}}+{tg}_{2_{i}}\right)\left(1-\dfrac{T_{4_{i}}}{T_{5_{i}}}\right)-{tg}_{3_{i}}\right),
\end{array}
\right.
\end{eqnarray}\allowdisplaybreaks
where $i$ is the generator index. For generator $i$,  the terminal voltage phasor $E_{t_{i}}=e_{R_{i}}+je_{I_{i}}$ and the terminal current phasor $I_{t_{i}}=i_{R_{i}}+ji_{I_{i}}$ can be measured and used as outputs from actual PMU measurements. The physical meaning of all the parameters in~\eqref{gen model} are included in the nomenclature section.

\begin{rem}\normalfont 
	For the above power system model, we treat the exciter and governor control system variables as state variables and thus there are no control inputs in the system model. 
\end{rem}
The $T_{m_{i}}$, $T_{e_{i}}$, $i_{d_{i}}$, $i_{q_{i}}$ $V_{A_{i}}$, $S_{E_{i}}$, and $D_i$ in (\ref{gen model})
can be written as functions of the states:
\allowdisplaybreaks
\setcounter{equation}{1}
\begin{subequations} \label{temp}
	\begin{align}
		& T_{m_{i}} = \dfrac{T_{4_{i}}}{T_{5_{i}}}\left(\dfrac{T_{3_{i}}}{T_{c_{i}}}{tg}_{1_{i}} + {tg}_{2_{i}}\right) + {tg}_{3_{i}} \\
		& \Psi_{R_{i}} = e'_{di}\sin\delta_i+e'_{q_{i}}\cos\delta_i \\
		& \Psi_{I_{i}} = e'_{q_{i}}\sin\delta_i-e'_{di}\cos\delta_i \\
		& I_{t_{i}} = \boldsymbol{\overline{Y}}_i(\m \Psi_{R}+j \m \Psi_{I}) \\
		& i_{R_{i}} = \operatorname{Re}(I_{t_{i}}) \\
		& i_{I_{i}} = \operatorname{Im}(I_{t_{i}}) \\
		& i_{q_{i}} = \dfrac{S_B}{S_{N_{i}}}(i_{I_{i}}\sin\delta_i+i_{R_{i}}\cos\delta_i) \\
		& i_{d_{i}} = \dfrac{S_B}{S_{N_{i}}}(i_{R_{i}}\sin\delta_i-i_{I_{i}}\cos\delta_i) \\
		& e_{q_{i}} = e'_{q_{i}}-x'_{d_{i}}i_{d_{i}} \\
		& e_{d_{i}} = e'_{di}+x'_{q_{i}} i_{q_{i}} \\
		& P_{e_{i}} = e_{q_{i}}i_{q_{i}}+e_{d_{i}}i_{d_{i}} \\
		& T_{e_{i}} = \dfrac{S_B}{S_{N_{i}}} P_{e_{i}} \\
		& V_{FB_{i}}=\dfrac{K_{F_{i}}}{T_{F_{i}}}(E_{fd_{i}}-R_{f_{i}}) \\
		& V_{TR_{i}}=\sqrt{{e_{q_{i}}}^2+{e_{d_{i}}}^2} \\
		& V_{A_{i}}= -V_{FB_{i}} - V_{TR_{i}} + exc_i^3 \\
		& S_{E_{i}}= exc_i^1\, e^{exc_i^2 |E_{fd_{i}}|} \textrm{sgn}(E_{fd_{i}}) \\
		& \omega_{e_{i}} = \dfrac{1}{\omega_0}(\omega_{f_{i}} - \omega_i) \\
		& d_i = P_{m_{i}}^0 + \dfrac{1}{r_i}\,\omega_{e_{i}} \\
		& D_i =  \left\{
		\begin{array}{ll}
			0, & \quad d_i \le 0 \\ 
			d_i, & \quad 0 < d_i \le T^{\max}_{i} \\
			T^{\max}_i, & \quad d_i > T^{\max}_i. \\
		\end{array} \right.
	\end{align}
\end{subequations}
The state vector $\boldsymbol{x}$ and output vector $\boldsymbol{y}$ are
\begin{subequations}
	\begin{align}
		\boldsymbol{x} = & \bmat{\boldsymbol{\delta}^{\top} \, \boldsymbol{\omega}^{\top} \, \boldsymbol{e'_{q}}^{\top} \, \boldsymbol{e'_{d}}^{\top} \, \boldsymbol{V_R}^{\top} \, \boldsymbol{E_{fd}}^{\top} \, \boldsymbol{R_f}^{\top} \, \boldsymbol{{tg}_{1}}^{\top} \, \boldsymbol{{tg}_{2}}^{\top} \, \boldsymbol{{tg}_{3}}^{\top}}^{\top} \notag \\ 
		\boldsymbol{y} = &\bmat{\boldsymbol{e_R}^{\top} \quad \boldsymbol{e_I}^{\top} \quad \boldsymbol{i_R}^{\top} \quad \boldsymbol{i_I}^{\top}}^{\top}, \notag
	\end{align}
\end{subequations}
and the power system dynamics can be written as:
\setcounter{equation}{2}
\begin{equation}~\label{equ:NLD}
\left\{
\arraycolsep=1.4pt\def\arraystretch{1.2}
\begin{array}{ll}
\dot{\m x}(t)&=\m f(\m x) \\
\m y(t) &= \m h(\m x).
\end{array}
\right.
\end{equation}
In \eqref{temp} the outputs $i_{R_{i}}$ and $i_{I_{i}}$ are written as functions of $\boldsymbol{x}$.
Similarly, the outputs $e_{R_{i}}$ and $e_{I_{i}}$ can also be written as functions of $\boldsymbol{x}$:
\setcounter{equation}{3}
\begin{subequations}
	\begin{align}
		e_{R_{i}}& = e_{d_{i}}\sin\delta_i+e_{q_{i}}\cos\delta_i \\
		e_{I_{i}} &= e_{q_{i}}\sin\delta_i-e_{d_{i}}\cos\delta_i.
	\end{align}
\end{subequations}
\subsection{Linearized Power System Model}~\label{sec:LinearizedDyn}
For a large-scale power system, the nonlinear model can be difficult to analyze, necessitating a simpler, linear time-invariant (LTI) representation of the system~\cite{Chakra2013}. 
The power system dynamics can be linearized by considering a small perturbation over an existing equilibrium point. 
The following assumption is needed to construct the small-signal, linearized model of the nonlinear power system.
\begin{assumption}\normalfont 
	For the nonlinear dynamical system in (\ref{gen model}) 
	there exists an equilibrium point denoted as:
	$$\m x^{*\top}=\bmat{\boldsymbol{\delta}^{\top} \, \boldsymbol{\omega}^{\top} \, \boldsymbol{e'_{q}}^{\top} \, \boldsymbol{e'_{d}}^{\top} \, \boldsymbol{V_R}^{\top} \, \boldsymbol{E_{fd}}^{\top} \, \boldsymbol{R_f}^{\top} \, \boldsymbol{{tg}_{1}}^{\top} \, \boldsymbol{{tg}_{2}}^{\top} \, \boldsymbol{{tg}_{3}}^{\top}}^{*}\hspace{-0.1cm}.$$
\end{assumption}
The above assumption is typical in transient analysis studies for power systems and other engineering applications 
modeled by nonlinear DAEs~\cite{Hill1989}. 
Denote by $\tilde{\m x} \in \mathbb{R}^{10\,n_g}$ the deviations of the state from the equilibrium point 
and $\tilde{\m y}_q  \in \mathbb{R}^{4q}$ the deviations of the outputs from the outputs at the equilibrium point. 
The small-signal dynamics can be written as:
\begin{equation}~\label{equ:LTICompact}
\left\{
\arraycolsep=1.4pt\def\arraystretch{1.2}
\begin{array}{ll}
\dot{\tilde{\m x}}(t) &= \mA \, \tilde{\m x}(t) \\
\tilde{\m y}_q(t) &= \mC_q\, \tilde{\m x}(t).
\end{array}
\right.
\end{equation}
where the system matrix $\mA \in \mathbb{R}^{10n_g \times 10n_g}$ is defined by the parameters of the generators, loads, transmission lines, and the topology of the power network, and $\mC_q \in \mathbb{R}^{4q\times 10\,n_g}$ depends on the specific PMU placement, where $q$ is the total number of  PMUs with four measurements each. In what follows, we use the notations $\m x$ and $\m y_q$ instead of $\tilde{\m x}$ and $\tilde{\m y}_q$ for simplicity.

\allowdisplaybreaks

\section{Unknown Inputs \& Attack-Threat Model:\\ The Physical Meaning}~\label{sec:UIAttackModel}


Although the modeling of the power system dynamics has been the subject of extensive research studies, a gap still exists between our mathematical understanding of the power system physics and the actual dynamic processes. Therefore, assuming that the developed dynamical models are \textit{perfectly accurate} can generate sub-optimal control or estimation laws. Consequently, various control and estimation theory studies have investigated methods that address the aforementioned discrepancy between the models and the actual physics---for power systems and other dynamical systems.

Here, we discuss how these discrepancies can be systematically incorporated into the power system dynamics and present physical interpretations of UIs and potential CAs---exemplifying these discrepancies. In this paper, and by definition, we consider UIs, denoted by $\m w(t)$, and CAs, denoted by $\m v_q(t)$, to be unknown quantities that affect the process dynamics and PMU output measurements, respectively.

\subsection{Modeling Unknown Inputs}
The nominal system dynamics for a controlled power system can be given by $\dot{\m x}(t)=\m f(\m x, \m u)=\mA \m x(t)+\mB_u \m u(t).$
\begin{rem} \normalfont
For the $10$th order power system model the controls, $\m u(t)$, are incorporated with the power system dynamics and states---we consider that the controls have a dynamic model as well. In that case, $\mB_u$ and $\m u(t)$ are both zeroes, unless there are other power system controls to be considered. 
\end{rem}
Here, we consider the nominal system dynamics to be a function of $\m w(t)$, or $\dot{\m x}(t)=\tilde{\m f}(\m x,\m u, \m w)$. For power systems, the UIs affecting the system dynamics can include $\m u_d$~(representing the unknown plant disturbances), $\m u_u$ (denoting the unknown control inputs), and $\m u_a$~(depicting potential actuator faults). For simplicity, we can combine $\m u_d, \m u_u, \m u_a$ into one UI quantity, $\m w(t)$, defined as $\m w(t) = \bmat{\m u_d^{\top}(t) & \m u_u^{\top}(t) & \m u_a^{\top}(t)}^{\top} \in \mathbb{R}^{n_w },$
and then write the process dynamics under UIs as
\begin{equation}
\dot{\m x}(t) = \tilde{\m f}(\m x,\m u,\m w) = \mA \m x(t) + \mB_u  \m u(t)+ \mB_w  \m w(t),
\end{equation}
where $\mB_w$ is a known weight distribution matrix that defines the distribution of UIs with respect to each state equation $\dot{x}_i$. For the dynamical system in (\ref{gen model}), matrix $\mB_w \in \mathbb{R}^{10n_g \times n_w}$. The term $\mB_w\m w(t)$ models a general class of UIs such as uncertainties related to variable loads, nonlinearities, modeling uncertainties and unknown parameters, noise, parameter variations, unmeasurable system inputs, model reduction errors, and actuator faults~\cite{Chen2012,Pertew2005}.  

For example, the equation $\dot{x}_1=\dot{\delta}_1=x_2 -\omega_0 = \omega_1 - \omega_0$ most likely has no UIs, as there is no modeling uncertainty related to that process. Also, actuator faults on that equation are not inconceivable. 
Hence, the first row of $\m B_w$ can be identically zero. Furthermore, if one of the parameters in~\eqref{gen model} are unknown, this unknown parameter can be augmented to $\m w(t)$. Furthermore, the unknown inputs we are considering are influencing all the buses in the power system. 
Precisely, for any state-evolution $x_i(t)$, we have:
$\dot{x}_i(t)=\mA_{i} \m x(t) + \mB_{w_{i}} \m w(t)=\mA_{i} \m x(t) + \sum_{j=1}^{n_w} B_{w_{ij}} w_j(t),$
$\forall i =1,2,\ldots, n,$ where  $\mA_{i}$ is the $i$th row of the $\mA$-matrix and $\mB_{i}$ is the $i$th row of the $\mB_w$-matrix, which entails that each bus can be potentially influenced by a combination of UIs. Hence, even if we have so many variations, these variations are causing disturbances to all the states in the power system. 
\begin{rem} \normalfont
Note that for a large-scale system it can be a daunting task to determine $\mB_w$. Hence, state estimators should ideally consider worst case scenarios with UIs, process noise, and measurement noise. As a result, assuming a random $\mB_w$ matrix and then designing an estimator based on that would consequently lead to a more robust estimator/observer design. \end{rem}
\subsection{Modeling Cyber Attacks}
As mentioned in the introduction, NESCOR developed cyber-security failure scenarios with corresponding impact analyses~\cite{NESCOR2014}.	The report classifies potential failure scenarios into different classes, including wide area monitoring, protection, and control (WAMPAC)---this paper's main focus. 
Here, and relevant to the physical meaning of CAs (or attack-vectors), we define $\m v_q(t)\in \mathbb{R}^{4q}$ as a CA that is a function of time, used to depict the aforementioned WAMPAC failure scenarios. Note that many entries in this vector are zero as an attacker might not have the ability to attack all measurements simultaneously. Under a wide class of attacks, the output measurement equation can be written as
		\begin{equation}~\label{equ:attackedOutput}
		\m y_q(t)=\mC_q \m x(t) + \m v_q(t).
		\end{equation}
		In this formulation, $\m v_q(t)$ can encapsulate a CA with three different physical meanings and classes:
		\begin{enumerate}
			\item The first class is the \textit{data integrity attacks} where an attacker attempts to change the output measurement of a sensor; see~\cite{Sridhar2010,Giani2013} for recent results on data integrity attacks in smart grids. 
			\item  The second class is the replay attacks---the attacker replays a previous snapshot of a valid
			data packet~\cite{Tran2013}.
			\item The third class is the \textit{denial of service (DoS)} attack---the attacker introduces a denial in communication of the measurement. The authors in~\cite{Liu2013} discuss DoS attacks on load frequency control in smart grid.
		\end{enumerate}  
		While we define $\m v_q(t)$ in this paper to be a \textit{cyber-attack} or \textit{attack vector}, this definition encapsulate the above classes of ``attacks". Furthermore, another physical meaning for $\m v_q(t)$ is bad data. Bad data occurs when (1) a redundant measurement
		is erroneous, which can be detected by statistical tests based on measurement residuals, (2) observations may be corrupted with abnormally large measurement errors, (3) large unexpected meter and communication errors, or (4) malfunctioning sensors; see ~\cite{Chen2006} for bad data definitions. Hence, bad data can be different from CAs---CAs attempt to adversely influence the estimates. However, both (bad data attacks and CAs) can lead to negative consequences and can share \textit{mathematically equivalent meaning with varying threat levels.}
\subsection{Attacker's Objective}
An attacker wants to cause significant changes to the transmitted PMU data. Since this data is bound to be used for real-time control in smart grids (see U.S. Department of Energy (DOE) and NASPI mandates~\cite{DOE2014,NASPI,NESCOR2014}), a change in these estimates/measurements can cause significant alterations to the corresponding feedback control signals. In fact, the executive summary from a recently published U.S. DOE report highlights the inevitable usability of PMU measurements\footnote{From a DOE report: \textit{The Western Electricity Coordinating Council has determined that it can increase the energy flow along the California-Oregon Intertie by 100 MW or more using synchrophasor data for real-time control---reducing energy costs by an estimated \$35 million to \$75 million over 40 years without any new high-voltage capital investments \cite{DOE2014}}.}.
\section{State Estimation Method Under UIs and CAs}~\label{sec:StateEstimation}
With the integration of PMUs, an observer or a DSE method can be utilized to estimate the internal state of the generators. Observers can be viewed as computer programs running online simulations---they can be easily programmed and integrated into control centers. Observers differ from KF-based estimators in the sense that no assumptions are made on the distribution of measurement and process noise, i.e., statistical information related to noise distribution is not available. The objective of this section is to investigate a robust observer for power systems with real-time PMU measurements. 
\subsection{Sliding-Mode Observer for Power Systems}~\label{sec:ZakSMO}
A variable structure control or sliding model control is a nonlinear control method whose structure depends on the current state of the system. Similar to sliding mode controllers, sliding mode observers~(SMO) are nonlinear observers that possess the ability to drive the estimation error, the difference between the actual and estimated states, to zero or to a bounded neighborhood in finite time. Similar to some Kalman filter-based methods, SMOs have high resilience to measurement noise. In~\cite{Utkin2009}, approaches for effective sliding mode control in electro-mechanical systems are discussed. Here we present a succinct representation of the SMO architecture developed in~\cite{Zak2005}. For simplicity, we use $\m x$ as the state vector of the linearized power system, rather than $\tilde{\m x}$ and $\m y$ as the outputs from PMUs, rather than $\tilde{\m y}$. 
As discussed in previous sections, the linearized power system dynamics under UIs and CAs can be written as
\begin{align}~\label{equ:plantDynamics}
\left\{
\arraycolsep=1.4pt\def\arraystretch{1.2}
\begin{array}{lll}
\dot{\m x}(t) &=&  \mA \m x(t) + \mB_w  \m w(t) \\
\m y_q(t) &=& \mC_q \m x(t)+\m v_q(t),
\end{array}
\right.
\end{align}
where for the system described in (\ref{equ:NLD}) there are $10\,n_g$ states, $n_w$ unknown plant inputs, and $4\,q$ measurements.
\begin{assumption}~\label{ass:Obs}\normalfont
The above dynamical system is said to be observable if the observability matrix $\mathcal{O}$, defined as
$\mathcal{O} = \bmat{
\mC_q^{\top} &(\mC_q\mA)^{\top}&
\cdots &
(\mC_q\mA^{10n_g-1})^{\top}
}^{\top}$
has full rank. The full-rank condition on the system implies that a matrix $\mL_q \in \mathbb{R}^{10n_g \times 4q}$ can be found such that matrix $(\mA-\mL_q\mC_q)$ is asymptotically stable with eigenvalues having strictly negative real parts. While this assumption might be very restrictive, it is not a necessary condition for the estimator we discuss next. This assumption is relaxed to the detectability of the pair $(\mA,\mC_q)$. The power system is detectable if all the unstable modes are observable---verified via the PBH test:
$$ \rank\bmat{\lambda_i \mI - \mA \\ \mC_q} = 10n_g, \, \forall \,  \lambda_i > 0,$$
where $\lambda_i$ belongs to set of eigenvalues of $\mA$. Also, we the observer rank-matching condition is satisfied, that is: 
$ \rank(\mC_q\mB_w)=\rank(\mB_w)=\zeta.$
\end{assumption}
The objective of an observer design is to drive the estimation error to zero within a reasonable amount of time. Accurate state estimates can be utilized to design local or global state feedback control laws, steering the system response towards a desirable behavior. Let $\hat{\m x}(t)$ and $\m e(t)=\m x(t) - \hat{\m x}(t)$ denote the estimated states and the estimation error. 
\subsection{SMO Dynamics \& Design Algorithm}
The SMO for the linearized power system dynamics~\eqref{equ:plantDynamics} can be written as
\begin{eqnarray}~\label{equ:SMO}
\left\{
\arraycolsep=1.4pt\def\arraystretch{1.2}
\begin{array}{lll}
\dot{\hat{\m x}}(t) &= \mA \hat{\m x}(t)+\hspace{-0.1cm} \mL_q(\m y_q(t)  - \hspace{-0.1cm}\hat{\m y}_q(t)) -\mB_w\mE(\hat{\m y}_q,\m y_q,\eta)  \\
\hat{\m y}_q(t) &= \mC_q \hat{\m x}(t),
\end{array}
\right.
\end{eqnarray}
where 
$\m y_q$ is readily available signals for the observer, and $\mE(\cdot)$ is defined as
\begin{equation*}
\mE(\cdot) = \begin{cases}
\eta \dfrac{\mF_q(\hat{\m y}_q-\m y_q)}{\normof{\mF_q(\hat{\m y}_q-\m y_q)}_2+\nu} &\text{if $\,\,\mF_q(\hat{\m y}_q-\m y_q) \neq \textbf{0}$}\\
\textbf{0}  &\text{if $\,\,\mF_q(\hat{\m y}_q-\m y_q) = \textbf{0}$},
\end{cases}
\end{equation*}
where:
\begin{itemize}
\item $\eta > 1$ is the SMO gain and $\nu$ is a smoothing parameter (small positive number),
\item $\mF_q \in \mathbb{R}^{n_w \times 4q}$ satisfies the following matrix equality $$\boxed{\mF_q\mC_q=\mB_w^{\top}\mP,}$$
\item $\mL_q \in \mathbb{R}^{10n_g \times 4q}$ is chosen to guarantee the asymptotic stability of $\mA-\mL_q\mC_q$.
\end{itemize} 
 Hence, for any positive definite symmetric matrix $\mQ$, there is a unique symmetric positive definite matrix $\mP\in \mathbb{R}^{10n_g \times 10n_g}$ such that $\mP$ satisfies the Lyapunov matrix equation,
$$\boxed{(\mA-\mL_q\mC_q)^{\top}\mP+\mP(\mA-\mL_q\mC_q) = -\mQ, \,\,\, \mP=\mP^{\top} \succ 0.}$$
The nonlinear vector function, $\mE(\cdot)$, guarantees that the estimation error is insensitive to the UI $\m w(t)$ and the estimation error converges asymptotically to zero. If for the chosen $\mQ$, no matrix $\mF_q$ satisfies the above equality, another matrix $\mQ$ can be chosen. Note that the SMO can deal with a wide range of unknown parameters and inputs (affecting states evolution), yet it cannot tolerate a severe CA against the PMU measurements. In this paper, the framework we develop addresses this limitation through the dynamic risk mitigation algorithm that utilizes CAs estimation and a detection filter (Sections~\ref{sec:AttackEst} and \ref{sec:AttackMitig}).

A design algorithm for the aforementioned SMO can be found in~\cite{Zak2005}, which presents a systematic way of obtaining the gain matrices for reduced-order observers. Here, we present a simple solution to the observer design problem. 

The above boxed equations are the main matrix-equalities needed to solve for the observer matrices $\mF_q,\mP,$ and $\mL_q$---guaranteeing the asymptotic stability of the estimation error, and the convergence of the state-estimates to actual ones. However, these equalities are bi-linear matrix equalities, due to the presence of the $\mP\mL_q\mC_q$ term in the Lyapunov matrix equation. Using the LMI trick by setting $\mY=\mP\mL_q$, we can rewrite the above system of linear matrix equations as:
	 	\boxalign{\begin{align}
		 		\mA^{\top}\mP+\mP\mA -\mC_q^{\top}\mY^{\top}-\mY\mC_q& = -\mQ \nonumber\\
		 	\mP&=\mP^{\top} \label{equ:LMIDesign}\\
		 	\mF_q\mC_q &=\mB_w^{\top}\mP\nonumber .
		 	\end{align}	 }	
After obtaining $\mP,\mF_q,\mY$, and computing\footnote{The computation of these matrices is performed offline, i.e., the observer is designed apriori. In Section~\ref{sec:sim}, we present the number of free and linear variables, as well as the offline running time of the observer design problem for the considered power system.} $\mL_q=\mP^{-1}\mY$, the SMO can be implemented via a numerical simulation. The above system of equations can be easily solved via any semidefinite program solvers such as \texttt{CVX}~\cite{CVX1,CVX2}, \texttt{YALMIP}~\cite{YALMIP}, or MATLAB's \texttt{LMI} solver.

\section{Asymptotic Reconstruction of UIs \& CAs}~\label{sec:AttackEst}
In last section, we introduce real-time observers for a power system. 
Here, we present estimation methods for the vectors of UIs, $\m w(t)$, and potential attacks, $\m v_q(t)$. To our knowledge, this approach has never been utilized in power systems with observers. This approach we discuss here, however, does not provide strict guarantees on the convergence of the estimates of these quantities, yet it is significant in the developed risk mitigation strategy. To guarantee the detection of CAs and compromised PMU measurements, we also discuss an attack detection algorithm with performance guarantees.

\subsection{Estimating Unknown Inputs} 

As discussed earlier, the designed SMO guarantees the asymptotic convergence of the state estimates to the actual ones. Substituting the differential equations governing the dynamics of the power system~\eqref{equ:plantDynamics} and the SMO~\eqref{equ:SMO} into the estimation error dynamics, we obtain
\begin{eqnarray}
\dot{\m e}(t) & = & \dot{\m x}(t)-\dot{\hat{\m x}}(t) \\
& = & \left(\mA-\mL_q\mC_q\right) (\m x(t)-\hat{\m x}(t))  +\mB_w\m w(t) \nonumber \\
& &-\mB_w\mE(\hat{\m y}_q,\m y_q,\eta) \nonumber \\
& = & \left(\mA-\mL_q\mC_q\right) \m e(t) +\mB_w\m w(t)-\mB_w\mE(\m e, \eta). \label{equ:errordynamics} 
\end{eqnarray}
This SMO is designed to guarantee that $\hat{\m x}(t)$ is the asymptotic estimate of $\m x(t)$. Since it is assumed that $\mB_w$ is a full-rank matrix, the following UI approximation holds:
\begin{equation}~\label{equ:u2estimate}
\hat{\m w}(t) \approx \mE(\hat{\m y}_q(t),\m y_q(t),\eta).
\end{equation}
The above estimates, as reported in~\cite{Kalsi2011}, requires further low-pass filtering which can be very heuristic. Here, we suggest an alternative to the UI estimation assuming that the state estimates converge to the actual ones asymptotically.

First, we write the discretized version of the power system dynamics:
$$ \m x(k+1)=\tilde{\mA}\m x(k) + \tilde{\mB}_u \m u(k) + \tilde{\mB}_w \m w(k),$$
where $\tilde{\mA}=e^{\mA h}$, $\tilde{\mB}_{u}=\int_{0}^{h}e^{\mA\tau}\mB_{u}\,d\tau$, and  $\tilde{\mB}_{w}=\int_{0}^{h}e^{\mA\tau}\mB_{w}\,d\tau$ are the discrete version of the state-space matrices. Since the observer design guarantees the convergence of the state estimates, $\hat{\m x}(t)$ or $\hat{\m x}(k)$, and $\hat{\m x}(k)$ is available for all $k$, then the vector of UI $\m w(k)$ can be approximated as follows. Substituting $\m x(k)$ by $\hat{\m x}(k)$ in the discretized dynamics of the power system, we obtain:
$$ \hat{\m x}(k+1)=\tilde{\mA}\hat{\m x}(k) + \tilde{\mB}_u \m u(k) + \tilde{\mB}_w \hat{\m w}(k).$$ 
Then, another estimate for the UI vector can be generated as:
\begin{equation}~\label{equ:westimate}
\boxed{\hat{\m w}(k) = \left(\tilde{\m B}_w\right)^{\dagger}\left(\hat{\m x}(k+1)-\tilde{\mA}\hat{\m x}(k) -\tilde{\mB}_u \m u(k)\right),}
\end{equation}
assuming that $\tilde{\m B}_w$ has full column rank and its left pseudo-inverse exists. Note that this estimation of the UI vector uses the generated estimates of one subsequent time period ($\hat{\m x}(k+1)$) and the actual control (if the latter exists in the model). 
This assumption is not restricting as observers/estimators are computer programs that run in parallel with the plants or dynamic processes.

\subsection{Estimating CAs} 

Attacks against synchrophasor measurements can be modeled in various scenarios. One possible scenario is the injection of malicious signals that alter the values of the measurements in the data packets sent from PMUs to PDCs and control centers, in addition to PMUs malfunctions. As in~\eqref{equ:plantDynamics},  a real-time CA $\m v_q(t)$  is included to alter the PMU measurements. 
The objective of this section is to apply an attack detection technique based on the estimation of CAs. Assuming an identical SMO architecture as the one presented in the previous section, an estimate of the CA, $\hat{\m v}_q(t)$, is derived in~\cite{Kalsi2011} and its dynamics takes the following form:
\begin{equation}~\label{equ:vqestimate}
\hat{\m v}_q(t)=-(\mF_q\mC_q\mL_q)^{\dagger}(\mF_q\mC_q\mB_w)(\bar{\mE}(t)-\hat{\m w}(t))\vspace{-0.2cm}
\end{equation}
$$+(\mF_q\mC_q\mL_q)^{\dagger}\mF_q\dot{\hat{\m v}}_q(t),$$
where $\hat{\m w}(t)$ is given in~\eqref{equ:u2estimate}, $\mF_q$ and $\mL_q$ are SMO design parameters, $\bar{\mE}(t)$ is selected such that the system is \textit{in sliding mode} along $\mF\mC_q\m e(t)=0$. In~\cite{Kalsi2011}, the authors assume that $\dot{\hat{\m v}}_q(t)\approx \textbf{0}$, which might not be a reasonable assumption in our application since an CA can be designed such that $\dot{\m v}_q(t)\neq \textbf{0}$. Rearranging~\eqref{equ:vqestimate}, we obtain
\begin{equation}~\label{equ:veqestimate2}
\dot{\hat{\m v}}_q(t) = \mV_1^{-1}\hat{\m v}_q(t)+\mV_1^{-1}\mV_2{\m m}(t),
\end{equation}
where
\begin{eqnarray*}
\mV_1 & = & (\mF_q\mC_q\mL_q)^{\dagger} \mF_q \in \mathbb{R}^{4q \times 4q},\,\, {\m m}(t) = \bmat{\hat{\m w}^{\top}(t) & \bar{\mE}^{\top}(t)}^{\top} \\
\mV_2 & = & \bmat{(\mF_q\mC_q\mL_q)^{\dagger}(\mF_q\mC_q\mB_w) & -(\mF_q\mC_q\mL_q)^{\dagger}(\mF_q\mC_q\mB_w) } \\
& & \in \mathbb{R}^{4q \times (n_w +10n_g)}.
\end{eqnarray*}
Note that $\mV_1$ is invertible. A more accurate estimate for the CA can further be obtained as
\begin{equation*}
\boxed{\hat{\m v}_q(t) = e^{\mV_1^{-1}(t-t_0)}\hat{\m v}_q(t_0) + \int_{t_0}^{t} e^{\mV_1^{-1}(t-\varphi)}\mV_1^{-1}\mV_2{\m m}(\varphi)  \, d\varphi.}
\end{equation*}

\subsection{Attack Detection Filter}
While the CA estimates generated from the methods discussed above can instantly identify compromised measurements for few time instances after the detection, the attack can propagate and influence the estimation of other measurements. 
In the case of lower sampling rates or computational power, another attack detector can be used. 
In~\cite{Pasq2013}, a robust attack identification \textit{filter} is proposed to detect the compromised nodes for longer time periods. 
We tailor this filter to our dynamical representation of the power system, which is also a dynamical system and takes the following form:
\begin{eqnarray}
\left\{
\arraycolsep=1.4pt\def\arraystretch{1.2}
\begin{array}{lll}
\dot{\m l}(t) & = & (\mA + \mA\mC_q^{\top}\mC_q)\m l(t) + \mA\mC_q^{\top}\m y_q(t)~\label{equ:filter1}\\
\m r(t) & = &  \m y_q(t)-\mC_q \m l(t),
\end{array}
\right.
\end{eqnarray}
where $\m l(t) \in \mathbb{R}^{10n_g}$ is the state of the filter and $\m r(t)\in \mathbb{R}^{4q}$ is the residual vector that determines the compromised measurements---the reader is referred to~\cite{Pasq2013} for more details on the filter design. The initial state of the filter, $\m l(t_0)$, is by definition equal to the initial state of the plant $\m x(t_0)$. Since initial conditions might not be available, the SMO discussed in Section~\ref{sec:ZakSMO} is utilized to generate $\m x(t_0) \approx \hat{\m x}(t_0)$. Hence, the SMO is necessary for the detection of the attack, i.e., we assume that the SMO is utilized for an initial period of time when measurements are not compromised. 

After generating the converging estimates of the states and UIs, the filter~\eqref{equ:filter1} generates real-time residuals $\m r(t)$. These residuals are then compared with a threshold to determine the most \textit{infected/attacked} nodes. The residuals here are analogous to the estimates of the CAs, $\hat{\m v}_q(t)$, which we derive in the previous section. It is significantly crucial for the attack detection filter and the CA estimators to obtain online computations of the residuals and estimates---the attacked measurements might adversely influence the estimation as the attacks can propagate in many networks. 

The risk mitigation algorithm we develop in the next section utilizes $\m r(t)$, $\hat{\m v}_q(t)$, and $\hat{\m w}(t)$ to determine the authenticity of PMU measurements, and identify the to-be-diagnosed measurements, while guaranteeing the observability of the power system through available measurements.

\section{Risk Mitigation---A Dynamic Response Model}~\label{sec:AttackMitig}

Here, we formulate a risk mitigation strategy given estimates of measured and estimated outputs and reconstructed UIs and CAs. 
The formulation uniquely integrates dynamic state estimation, considering attacks and UIs, with a integer linear programming formulation. 
\subsection{Weighted Deterministic Threat Level Formulation}~\label{sec:WDTL}

We consider the measured and estimated PMU signals as an essential deterministic component in the decision making problem that decides which PMUs should be disconnected from the network for a period of time, while performing typical troubleshooting and diagnosis. 
\begin{mydef}\normalfont
Given a dynamic system simulation for $\tau \in \left[ kT,(k+1)T\right]$, where $T$ is any simulation time period, the weighted deterministic threat level~(WDTL) vector $\m z$ for all PMU measurements is defined as
\begin{eqnarray}
\m z = \displaystyle \int_{kT}^{(k+1)T}\left[ \mY\big( \m y_q(\tau) - \hat{\m y}_q(\tau)\big)^2  +  \mW \big(\hat{\m w}(\tau)\big)^2 \, \right] d\tau, ~\label{equ:wresid} 
\end{eqnarray}
where $\mY \in \mathbb{R}^{4q \times 4q}$ and $\mW \in \mathbb{R}^{4q \times n_w}$ are constant weight matrices that assign weights for the estimation error~($\m y_q - \hat{\m y_q}$) and UI approximation $\hat{\m w}$. Note that $\big(\hat{\m w}(\tau)\big)^2$ is equivalent to the square of individual entries. 
\end{mydef} 
The scalar quantity $\m z_i$, the $i$th WDTL, depicts the threat level present in the $i$th PMU signal.  Ideally, if $\m z_i$ is large  the associated PMU must be isolated until the attack is physically mitigated. The quantity $\big( \m y_q(\tau) - \hat{\m y}_q(\tau)\big)$ can be replaced with either $\hat{\m v}_q(t)$ or $\m r(t)$.
\subsection{Dynamic Risk Mitigation Optimization Problem} ~\label{sec:DRMOP}
Deactivating a PMU may lead to a failure in dynamic state estimation, as explained in the following Remark~\ref{rem:obs}. 
Hence, an optimization-based framework is proposed to solve the problem with occasionally conflicting objectives.
\begin{rem} \normalfont~\label{rem:obs}
Recall that to design a dynamic state estimator under UIs and CAs, the power system defined in~\eqref{equ:plantDynamics} should satisfy certain rank conditions on the state-space matrices. For example, for the SMO observer, the following condition has to be satisfied:
$$\rank (\mC_q\mB_w) = \rank (\mB_w) = \zeta,$$
in addition to the detectability condition (Assumption~\ref{ass:Obs}). 
Deactivating a PMU causes a change in the $\mC_q$ matrix and might render the observer design infeasible. 
\end{rem}
\begin{mydef}\normalfont
Let $\pi_i$ be a binary decision variable that determines the connectivity of the $i$th PMU measurement in the next time period (i.e., $\tau \in [kT,(k+1)T]$):
\begin{equation*}
	\pi_i = \begin{cases}
	0 &\text{$\leftrightarrow$ $\,\,\, z_i - \gamma_i  \geq 0$}\\
	1 &\text{$\leftrightarrow$ $\,\,\, z_i - \gamma_i < 0$}.
	\end{cases}
\end{equation*}
\end{mydef}
If the WDTL for the $i$th measurement is smaller than a certain threshold $\gamma_i$, the corresponding measurement qualifies to stay activated in the subsequent time period. This combinatorial condition can be represented as
\begin{eqnarray}
z_i - \gamma_i  + \pi_i M & \geq & 0 \\
z_i - \gamma_i  - (1-\pi_i) M & < & 0
\end{eqnarray}
where $M$ is a large positive constant~\cite{Morari1999}. We now formulate the dynamic risk mitigation optimization problem (DRMOP): 
\boxalign{\begin{eqnarray}
\maximize_{\m \pi} & &  \displaystyle \sum_{i=1}^{4q} \alpha_i \pi_i \label{equ:DRMOP}\\
\subjectto 
& & \pi_i = \{0,1\}, \forall i = \{1,2,\ldots 4q\}\\
 & & z_i - \gamma_i  + \pi_i M \geq  0  \,\,\, \\
& &  z_i - \gamma_i  - (1-\pi_i) M <  0 \,\,\, \\
& &  \displaystyle \sum_{i=1}^{4q} \beta_i \pi_i \leq Z \label{equ:DRMOPend} \\
& &\hspace{-0.1cm}\rank (\mC_q(\m \pi)\mB_w) = \zeta \label{equ:obs} \\
& & \hspace{-0.1cm}\rank\bmat{\lambda_i \mI - \mA \\ \mC_q(\m \pi)} = 10n_g,  \forall \,  \lambda_i > 0\label{equ:obs2}.
\end{eqnarray}}
To increase the observability of a power system, the formulated optimization problem---an integer linear program (ILP)---maximizes the weighted number of active PMU measurements in the next time period, finding the PMU measurements that have to be disabled for some period of time while ensuring the feasibility of dynamic state estimation. Albeit there are at most $q$ PMUs, we assume that there are $4\,q$ $\pi_i$'s. 

The first two constraints depict the logical representation of the binary variable $\pi_i$ in terms of the WDTL and the threshold. 
The third constraint represents a weight for each PMU. For example, if the $i$th PMU measurement is from a significantly important substation, the system operator can choose the corresponding weight $\beta_i$ to be greater than other weights. The two rank constraints~\eqref{equ:obs}--\eqref{equ:obs2} ensure that the dynamic state estimation formulated in the previous section is still feasible for the next time period; see Assumption~\ref{ass:Obs}.
Note that this problem is different from the optimal PMU placement problem~\cite{qi,adaptive}, in the sense that we already know the location of the PMUs. The DRMOP~\eqref{equ:DRMOP}--\eqref{equ:obs2} is a highly nonlinear, integer programming problem that cannot be solved efficiently---due to the two rank constraints. In the next section (Section~\ref{sec:DRMOPAlgo}), we present a dynamic risk mitigation solution algorithm by relaxing these assumptions. 
\subsection{Dynamic Risk Mitigation Algorithm}~\label{sec:DRMOPAlgo}
In Sections~\ref{sec:WDTL} and~\ref{sec:DRMOP}, we investigate two related problems for different time-scales: the estimation problem is executed in real time, whereas the DRMOP is solved after generating the estimates in the former problem. Here, we present an algorithm that jointly integrates these two problems, without including the rank constraints in the computation of the DRMOP solution, and hence guaranteeing fast solutions for the optimization problem.

\begin{algorithm}
\small	\caption{Dynamic Risk Mitigation Algorithm (DRMA)}\label{algo}
	\begin{algorithmic}[1]
\State \textbf{compute} small-signal system matrices $\mA,\mB_w,\mC_q$
\State \textbf{obtain} SMO matrices $\mL_q, \mF_q$ by solving~\eqref{equ:LMIDesign}
\State \textbf{formulate} the SMO dynamics as in~\eqref{equ:SMO} 
\State \textbf{set} $k:=0$
\State \textbf{for} $\tau \in [kT+\xi,(k+1)T]$ \label{step:aa}
\State \hspace{0.5cm} \textbf{measure} the PMU output $\m y(\tau)$
\State \hspace{0.5cm} \textbf{compute}~$\m r(\tau), \hat{\m y}_q(\tau),\hat{\m w}(\tau)$ from \eqref{equ:filter1}, \eqref{equ:SMO}, \eqref{equ:westimate} 
\State \hspace{0.5cm} \textbf{compute} WDTL $\m z$ from~\eqref{equ:wresid},  given $\mY,\mU$
\State \hspace{0.5cm} \textbf{solve} the DRMOP~\eqref{equ:DRMOP}--\eqref{equ:DRMOPend} for $\boldsymbol\pi=\bmat{\pi_1,\cdots,\pi_{4q}}$
\State \hspace{0.5cm} \textbf{update} $\mC_q=\mC_q(\boldsymbol\pi)$
\State \hspace{0.5cm} \textbf{if} \eqref{equ:obs} and \eqref{equ:obs2} are satisfied
\State \hspace{0.5cm} \hspace{0.5cm} \textbf{go} to Step~\ref{step:a}
\State \hspace{0.5cm} \textbf{else}
\State \hspace{0.5cm} \hspace{0.5cm} \textbf{solve} the DRMOP \eqref{equ:DRMOP}--\eqref{equ:DRMOPend} with relaxed conditions on some $\pi_i$'s and update $\mC_q$ 
\State \hspace{0.5cm}\textbf{end if}
\State \textbf{end for}
\State \textbf{set} $k:=k+1$; ~\label{step:a} \textbf{go} to Step~\ref{step:aa}
	\end{algorithmic}
\end{algorithm}

Algorithm~\ref{algo} illustrates the proposed dynamic risk mitigation algorithm. First, the small-signal matrices are computed given the nonlinear power system model\footnote{Note that for the $10$th order model, the controls are incorporated in the power system dynamics, and hence $\mB_u$ and $\m u(t)$ are zeros, yet the algorithm provided here is for the case when known controls are considered.}. 
The sliding-mode observer is then designed to ensure accurate state estimation, as in Section~\ref{sec:ZakSMO}. Since the rank-constraints are computationally challenging to be included in an integer linear programming, Algorithm~1 provides a solution to this constraint. 

Then, for $\tau \in [kT+\xi,(k+1)T]$, the quantities $\m r,\hat{\m y}_q$, and $\hat{\m w}$ are all computed. We assume that the computational time to solve the DRMOP~\eqref{equ:DRMOP}--\eqref{equ:DRMOPend} is $\xi$. After solving the ILP, the output matrix $\mC_q$ is updated, depending on the solution of the optimization problem, as the entries in the $\mC_q$ reflect the location of active and inactive PMUs. The matrix update might render the state estimation problem for $\tau \in [(k+1)T+\xi,(k+2)T]$ infeasible as the rank conditions might not be satisfied. To ensure that, these conditions can either be made a constraint in the optimization problem or a condition in the mitigation algorithm. If this rank conditions are not satisfied, some $\pi_i$'s can be reset and the DRMOP can be solved again. The counter $k$ is then incremented and the algorithm is applied for the following time periods. 
\begin{figure*}[!t]
	\centering
	\includegraphics[scale=0.75]{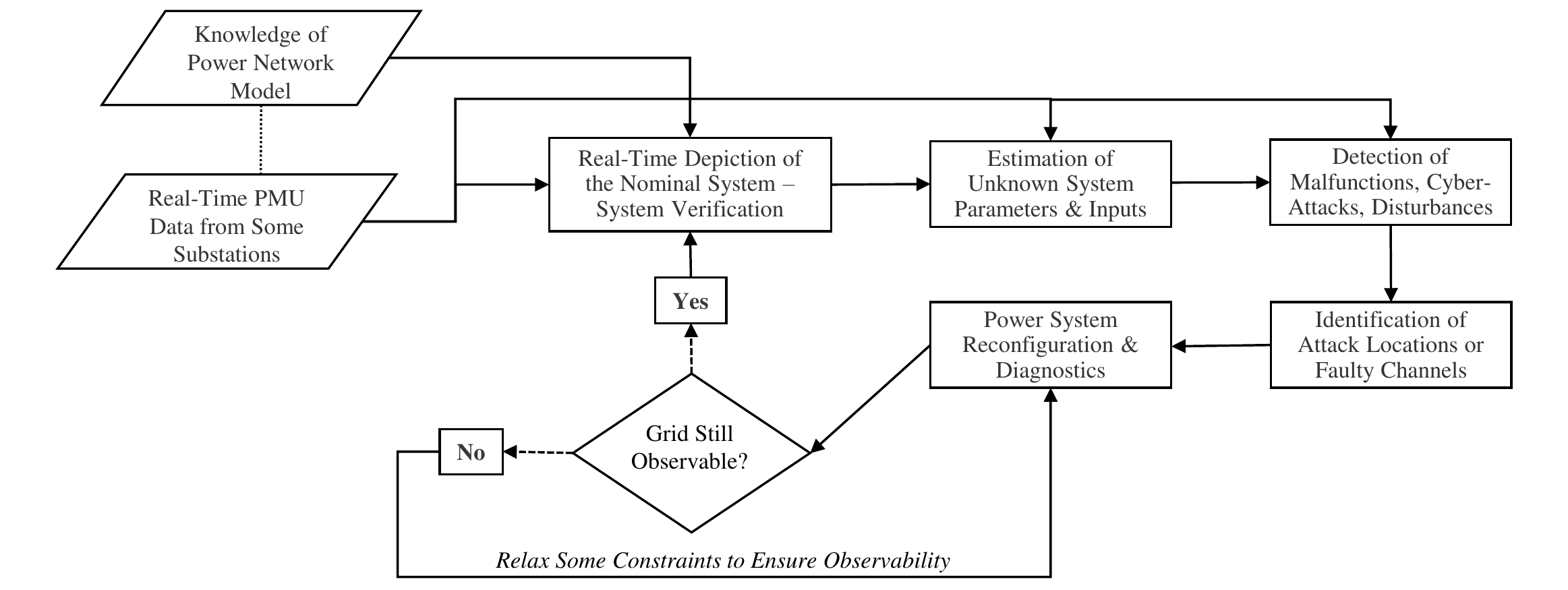}
	\caption{A flow chart depicting a high-level representation of the proposed risk mitigation strategy; see Section~\ref{sec:flowchart}.}
	\label{fig:flowchart}
\end{figure*}
\begin{rem} \normalfont
	For the developed algorithm, we assume that we are applying the observer from~\cite{KalsiDisser} to generate dynamic estimates, given that the power system is subjected to UIs and CAs. However, This assumption is not restricting. In fact, any other robust observer/estimator may be used for state estimation, and hence, the algorithm can be changed to reflect that update in the observer design. Subsequently, the matching rank condition can be replaced by other conditions that guarantee a fast reconstruction of state estimates.
\end{rem}\begin{rem} \normalfont
	The DRMOP assumes an initial power system \textit{configuration}, i.e., PMUs are placed in certain locations. Since the $\mC_q$ reflects the latter, the observer design would differ for various \textit{configurations} of PMUs. This influences the state and UI estimation, and hence the generation of the real-time weighted deterministic threat levels for all PMUs. Thus, the solution to the DRMOP varies for different PMU configurations, while guaranteeing the real-time observability of the power system through available measurements.
\end{rem}
\section{High-Level Solution Scheme for the Risk Mitigation Strategy---A Summary}~\label{sec:flowchart}
This section serves as a summary of the overall \textit{solution scheme} we develop in this paper. The high-level details of this scheme is illustrated in Fig.~\ref{fig:flowchart}. The presented solution scheme requires two \textit{essential} inputs: 
\begin{enumerate}
	\item[(a)] The potentially-incomplete knowledge of the power system model and parameters (Section~\ref{sec:SystemModel});
	\item[(b)] Real-time PMU measurements from a subset of the power network model (Section~\ref{sec:SystemModel}).
\end{enumerate}
Note that (a) and (b) are related in the sense that if the knowledge of a generator's parameters is available, it is possible to associate this knowledge to specific PMU measurements. 

Given these two inputs---(a) is static knowledge, while (b) is continuously updated---we construct a real-time depiction of the nominal system, i.e., the power system experiencing no CAs or major disturbances. This step is important as it verifies PMU measurements and the system model. Using the latter and real-time PMU data, we estimate unknown power system parameters and UIs (Section~\ref{sec:AttackEst}). Given that we have more accurate parameters, the detection of malfunctions, CAs, and major disturbances becomes possible; see Fig.~\ref{fig:flowchart}. 

However, the detection of a CA does not necessarily imply the knowledge of the source of attacks. Hence, the identification of PMU channels with faulty measurements is needed after the detection of such events (Section~\ref{sec:AttackMitig}). The faulty or attacked power system components are then diagnosed and reconfigured. The reconfiguration/diagnostics of the grid should, however, guarantee the observability of the grid~(Section \ref{sec:DRMOPAlgo}). After guaranteeing the latter, the power system is brought back to its initial nominal state. 
\section{Case Studies}~\label{sec:sim}
The developed methods are tested on a 16-machine, 68-bus system---extracted from Power System Toolbox (PST) \cite{chow1992toolbox}. This system is a reduced order, equivalent version of the interconnected New England test system and New York power system~\cite{Pal2013}.
The model discussed in Section~\ref{sec:SystemModel} is used and there are 160 state variables.
A total of $q=12$ PMUs are installed at the terminal bus of generators 1, 3, 4, 5, 6, 8, 9, 10, 12, 13, 15, and 16. 
Here this PMU placement is randomly chosen, while installing the PMUs at optimal locations to guarantee the best observability of the system dynamic sates is out of the scope of this paper. More details for that problem can be found in \cite{qi,adaptive}. 
The sampling rate of the measurements is set to be 60 frames per second to mimic the PMU sampling rate.

In the rest of this section, we present results for two scenarios. For Scenario I, dynamic state estimation only under UIs is performed and
an illustration on the estimation of UIs and states via the methods discussed in Section~\ref{sec:AttackEst} is provided. 
For Scenario II, we add CAs to some PMU measurements and show how the DRMOP can be utilized to estimate, detect, and filter out the presence of these attacks by leveraging the generated estimates from Scenario I. 
\subsection{Scenario I: Dynamic Reconstruction of UI \& DSE}
The objective of this section is to show the performance of the SMO in Section~\ref{sec:ZakSMO} in regards to the estimation of (a) the states of the 16 generators (160 states) and (b) UI reconstruction method (developed in Section~\ref{sec:AttackEst}). 
We perform dynamic state estimation over a time period of 20 seconds. 
We consider this experiment as a baseline for the Scenario II. The simultaneous estimation of states and UIs can be then utilized to determine the generators that are subject to the most disturbances through available PMU measurements. After the estimation of states and UIs, these quantities are then used to detect a CA against PMU measurements.

\textbf{Arbitrary Unknown Inputs} As discussed in Section~\ref{sec:UIAttackModel}, the UIs model a wide range of process uncertainties ranging from load deviations, bounded nonlinearities, and unmodeled dynamics, which can significantly influence the evolution of states due to their nature. However, UIs are not physically analogous to \textit{malicious} CAs, i.e., UIs exist due to phenomena related to the physics of the power system modeling.

For many dynamical systems, it can be hard to determine the impact of UIs. 
Hence, an ideal scenario would be to use different forms of time-varying UI functions, 
and a randomly generated $\mB_w$ matrix with significant magnitude. Here, we consider that the power system is subject to six different UI functions with different variations, magnitudes, and frequencies. 
The considered vector of UIs is as follows:
$$\small\m w(t)=\bmat{
w_1(t)&=&k_1\left(\cos(\psi_1 t)+e^{-2t}+\max\left(0,1-\frac{|t-5|}{3}\right)\right)\\
w_2(t)&=&k_1\sin(\psi_1 t)\\
w_3(t)&=&k_1\cos(\psi_1 t)\\
w_4(t)&=&k_2\sqq(\psi_2 t)\\
w_5(t)&=&k_2\sawtooth(\psi_2 t)\\
w_6(t)&=&k_2\left( \sin(\psi_2 t)+e^{-5t}\right)},$$
where $k_1,k_2$ and $\psi_1,\psi_2$ denote different magnitudes and frequencies of the UI signals, respectively. 
We choose $\psi_1=5, \psi_2=10$. To test the developed SMO and the UI estimator, we use small and large values for $k_1$ and $k_2$. 
Specifically, we choose two set of values for the magnitude as $k_1=0.01, k_2=0.02$ and $k_1=1, k_2=2$.

The $\mB_w \in \mathbb{R}^{160 \times 6}$ matrix is randomly chosen using the \texttt{randn} function in MATLAB. 
The Euclidean norm of $\mB_w$ is $\normof{\mB_w}= 13.8857$ which is significant in magnitude. 
Consequently, since $\mB_w$ is not sparse, the six chosen UIs influence the dynamics of the $160$ states of the power system, i.e., $\dot{x}_1(t)=\m a_1 \m x(t) + \sum_{i=1}^{6} B_{w_{1,i}} w_i(t)$, where $\m a_1$ is the first row of the $\mA$ matrix. The above UI setup is used in this experiment as an extreme scenario, as this allows to test the robustness of the estimator we develop in this paper. 
\begin{rem} \normalfont
Using large magnitudes for the UIs (i.e., large $k_1$ and $k_2$) leads to unrealistic behavior of the power system as each differential equation is adversely influenced by an unknown, exogenous quantity as described above. This scenario is less likely to occur in real applications, yet the result is included to show the robustness of the simultaneous estimation of the states and the proposed UI estimation scheme.
\end{rem} 

\textbf{SMO Design} After computing the linearized state-space matrices for the system ($\mA$ and $\mC_q$) and given $\mB_w$, we solve the LMIs in~\eqref{equ:LMIDesign} using CVX~\cite{CVX1} on Matlab. The SMO parameters are $\eta=8$ and $\nu=0.01$.
The numbers of linear and free variables involved in the semidefinite programming are $25760$ and $7968$\footnote{The number of linear and free variables in~\eqref{equ:LMIDesign} is equal to the number of entries of the symmetric positive definite matrix $\mP$ (linear vars.), and $\mL_q,\mF_q$. Since $\mP \in \mathbb{R}^{10n_g \times 10n_g}$ and $n_g=16$, the number of linear variables is $(160+1)(160)=25760$, while the number of free variables is equal to $n_w\cdot 4\,q + 10\,n_g\cdot 4\,q= 6\cdot 48+160\cdot 48=7968.$} with $13840$ constraints. The number of variables can be computed by counting the number of unique entries of the LMI in~\eqref{equ:LMIDesign}. 

The solution to this optimization problem is done offline, as most observer gain matrices are computed before the actual dynamic simulation. The simulations are performed on a 64-bit operating system, with Intel Core i7-4770S CPU 3.10GHz, equipped with 8 GB of RAM. The execution time for the offline SMO design~\eqref{equ:LMIDesign} is 5 minutes and 39 seconds (CVX converges after 42 iterations); see Remark~\ref{rem:smodesigntime} for a discussion on the running time of the offline SMO after the detection of attacks. The dynamic simulations for the power system and the observer dynamics are performed simultaneously using the \texttt{ode15s} solver with a computational time of nearly 6 seconds.

\begin{figure}[h]
	\centering
	\vspace{-0.31cm}\hspace{-0.2cm}
	\includegraphics[scale=0.33]{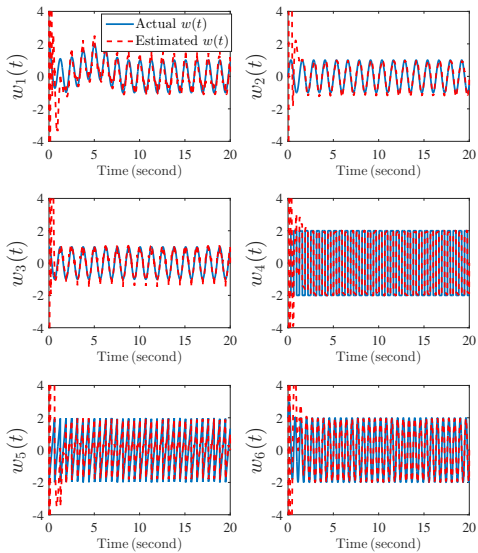}
	\vspace{-0.7cm}
	\caption{Converging estimation for the 6 UIs discussed above. The UI estimator successfully tracks the UIs for $k_1=1, k_2=2$ for different shapes of the UIs. The results for the UI reconstruction for $k_1=0.01, k_2=0.02$ are omitted; however, the results are similar to the case we present in this figure.}
	\label{fig:UI_Estimation}
\end{figure}

\textbf{State \& UI Reconstruction} After finding a solution for~\eqref{equ:LMIDesign}, we simulate the power system and generate estimates of the states $\m x(t)$ and the UIs $\m w(t)$ via the SMO design~\eqref{equ:SMO} and UI estimate~\eqref{equ:westimate}.
\begin{figure}[h]
	\centering
	\vspace{-0.21cm}	\includegraphics[scale=0.37]{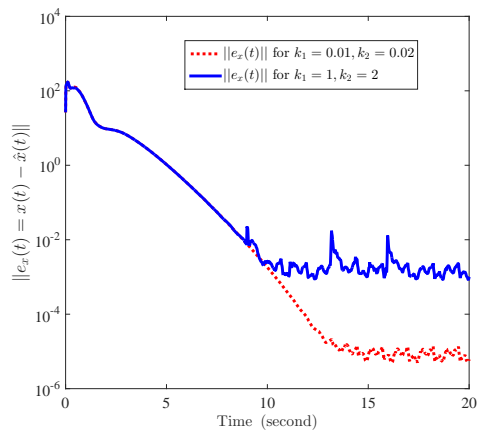}
	\vspace{-0.3cm}	\caption{Norm of the state estimation error for different magnitudes of UIs. 
		A logarithmic scale is used for y-axis, as initial values for $\normof{e_x(t)}_2$ are much higher than subsequent ones. 
		For larger magnitudes of UIs, the norm converges to a larger value, albeit it is still very small.}
	\label{fig:SENorm}
\end{figure}
In Fig.~\ref{fig:SENorm} we show the norm of the state estimation error for the above two sets of values of $k$'s. 
The estimation error norm is $ \normof{\m e_x(t)}_2 = \normof{\m x(t) -\hat{\m x}(t)}_2, \, \forall \, t \, \in [0,T]$, 
which indicates the performance of the SMO for all time instants and all generators. 
It is clearly seen that the estimation error converges to nearly zero---even for high magnitudes of UIs. 
This demonstrates that the state estimates for all generators are converging to the actual states of the power system. 
Moreover, Fig.~\ref{fig:UI_Estimation} shows the estimation of the six UIs given above with $k_1=1, k_2=2$. While the six UIs vary in terms of magnitude, frequency, and shape, the estimates generated by~\eqref{equ:westimate} are all very close to the actual UIs. 
\subsection{Scenario II: DSE Under UIs \& CAs} 
Here we present the case when some PMU measurements are compromised by a CA. 
The attacker's objective is to drastically alter the PMU measurements, thus influencing the decisions that could be made by the system operator. 
First, we discuss the attacker's strategy, i.e., what attack-signals are manipulating the measurements. Second, we present an algorithm that detects the presence of a CA and identifies the attacked measurement(s). Third, given the estimates of the attacks and state- and UI-estimation results, we apply the DRMOP and the DRMA. Finally, we show the impact of applying risk mitigation strategy on state estimation. 

\textbf{Artificial Cyber-Attacks} We present a hypothetical CA vector on four PMU measurements, which are the fifth to eighth measurements, i.e., $\bmat{{e_{R_{6}}}(t) & {e_{R_{8}}}(t) & {e_{R_{9}}}(t)& {e_{R_{10}}}(t)}^{\top}$. 
Note that these four measurements come from the PMUs installed at the terminal bus of Generators 6, 8, 9, and 10, respectively. 
Although we denote manipulation of some signals an \textit{attack}, this nomenclature is not restrictive; see Section~\ref{sec:UIAttackModel} and the NASPI report on faulty PMU measurements~\cite{NASPI}.
Since a total of $4\,q = 48$ measurements are available, the CA $\m v_q(t) \in \mathbb{R}^{48}$ can be constructed in terms of different unknown signal structures, as follows:
$$ \m v_q(t) = \bmat{\textbf{0}_{4} & \cos(t) & 2 \sawtooth(t) & 3 \sqq(t) & 4 \sin(t) & \textbf{0}_{40}}^{\top},$$
where the cosine, sawtooth, square, and sine signals are the actual attacks against the four PMU measurements with different magnitudes and variations.

\begin{figure}[h]
\centering
\hspace{-0.4cm}
	\includegraphics[scale=0.35]{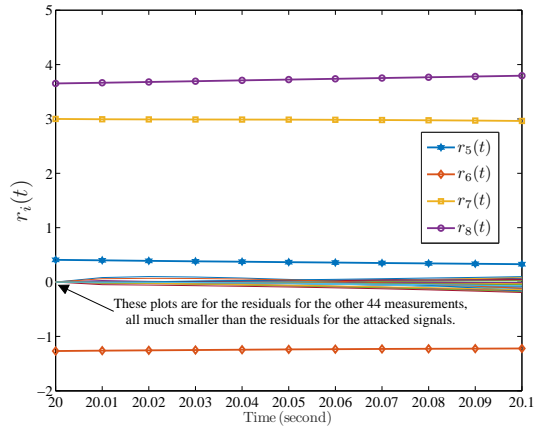}
	\caption{Residuals of the 48 measurements generated by the attack detection filter~\eqref{equ:filter1}. The residuals are notably similar to the actual attacks. 
	For example, for $t=20.05$s, $r_7(t)=2.986$, while $v_7(t)=3 \sqq(t) = 3$.} 
	\label{fig:res}
\end{figure}

\textbf{Attack Identification \& Residual Computation} Under the same UIs from Scenario I, a CA is artificially added after $t=20$s. Fig.~\ref{fig:res} shows the generation of residual vector, $\m r(t)$, from~\eqref{equ:filter1}.
It is seen that the residuals of measurements 5--8 with artificially added CAs are significantly higher than the other measurements without CAs. 

\textbf{Dynamic Risk Mitigation Algorithm} After designing the SMO for the power system, achieving desirable state and UI estimates (Scenario I), and generating residuals that are estimates of CAs, we simulate the DRMOP. We assume that all PMU measurements have the same weight in the objective function, i.e., $\alpha_i = 1, \,\, \forall i=1,\ldots,48$. The WDTL vector $\m z$ is computed for the 1-second time horizon (for $t=[20,21]$), and generic threshold is chosen as $\gamma=10$. Given $\m z(t)$ and the parameters of the DRMOP, the ILP is solved via YALMIP~\cite{YALMIP}. 
The optimal solution for the ILP yields $\boldsymbol\pi=\bmat{\textbf{1}_{4} & \textbf{0}_4 & \textbf{1}_{44}}^{\top}$, hence the PMU measurements 5--8 are the most \textit{infected} among the available 48 ones. This result confirms the findings of the attack detection filter in Fig.~\ref{fig:res}.

Following Algorithm~1, we check whether the solution generated by YALMIP violates the rank condition (Assumption~2). 
As measurements 5--8 are removed from the estimation process for diagnosis, the updated $\mC_q$ matrix, now a function of $\boldsymbol\pi$, is obtained---$\mC_q$ now has 44 rows instead of 48. The system is detectable and the rank-matching condition is still satisfied. Hence, no extra constraints should be reimposed on the ILP, as illustrated in Algorithm~1. After guaranteeing the necessary conditions on the existence of the dynamic state estimator, and updating $\mC_q$, simulations are performed again to regenerate the state estimates and weighted residual threat levels.

\begin{figure}[h]
	\centering	\vspace{-0.3cm}
	\includegraphics[scale=0.35]{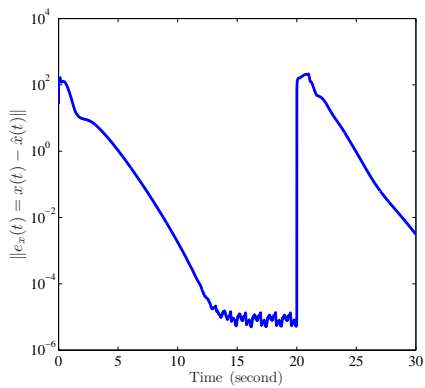}
	\caption{Norm of estimation error before, during, and after the CA is detected and isolated. For $20\,\textrm{s}\leq t \leq 21\,\textrm{s}$, the norm increases exponentially, signifying the occurrence of an attack or a
		significant disturbance. After the removal of the artificial attack due to the outcome of the DRMOP, the estimation error norm converges again to small values.
		}
	\label{fig:error_norm_attack}
\end{figure}

\begin{figure}[h]
\centering	\vspace{-0.51cm}	\hspace{-0.4cm}
	\includegraphics[scale=0.36]{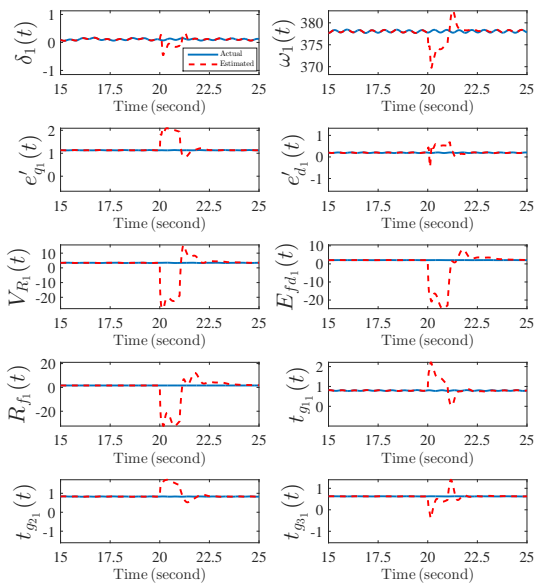}
	\vspace{-1cm}
	\caption{Estimation of the states of Generator 1 before, during, and after the detection and isolation of the CA. After the DRMA succeeds in detecting the compromised measurements and isolating them from the estimation process after $t=21$s, the state estimates converge again to the actual ones. }
	\label{fig:SEGen1_attack}
\end{figure}

\textbf{Post-Risk Mitigation} Fig.~\ref{fig:error_norm_attack} illustrates the impact of CAs on the state estimation process before, during, and after the attack is detected and isolated. 
Following the removal of the attacked measurements (and not the attack, as the attack cannot be physically controlled) at $t=21$s due to the risk mitigation strategy, the estimation error norm converges again to small values. 
Fig.~\ref{fig:SEGen1_attack} shows the impact of this strategy on DSE for Generator 1. 
During the short-lived CA, state estimates diverge. 
However, the risk mitigation strategy restores the estimates to their nominal status under UIs and CAs\footnote{While the CAs are still targeting the four PMU measurements after $t=21$s, the attacks become futile. Consequently, their impact on state estimation becomes nonexistent, as the four attacked measurements are isolated from the estimation process.}.
\begin{rem} \normalfont \label{rem:smodesigntime}
	The DRMA requires the redesign of the SMO immediately after the detection of the compromised PMU measurements. Since $\mC_q$ has less rows as the number of measurements are supposedly reduced after some of them are isolated, the SMO is designed again for an updated observer gain matrices $\mL_q$ and $\mF_q$. For a large scale system, the solution of the LMI in~\eqref{equ:LMIDesign} can take a significant amount of time. Hence, a database of the most possible PMU measurement configurations (different $\mC_q$'s) with corresponding SMO LMI solutions (different $\mL_q$'s and $\mF_q$'s) can be obtained offline, and stored when needed to guarantee a minimal off-time. 
\end{rem}
Note that for a different time period, the power system and the PMUs might encounter a different set of UIs or attack-vectors. Furthermore, the optimization problem can be redesigned to allow for the inclusion of the \textit{possibly,} \textit{now-safe} measurements. The optimal solution to the DRMOP is a trade-off between keeping the power system observable through the possible measurements---enabling state estimation and real-time monitoring---and guaranteeing that the system and the observer are robust to UIs and CAs. 

\subsection{Impact on Post-Estimation Process}
In power networks, many substations are not equipped with PMU devices. Hence, a robust DSE routine---that leverages measurements from neighboring nodes in a network---is necessary in generating state-estimates for the nodes without PMUs. As discussed earlier, a major purpose behind DSE is to use estimation of unknown, unmeasurable quantities for improved feedback control. In many cases, the estimates are used for control; in other cases, they are used for real-time depictions only. Here we discuss how erroneous, attacked state-estimation can influence power system stability if these estimates are used for control or re-dispatch purposes, such as the example given in Section~\ref{sec:UIAttackModel} from the DOE report~\cite{DOE2014}.

\begin{figure}[h]
\centering
	\includegraphics[scale=0.34]{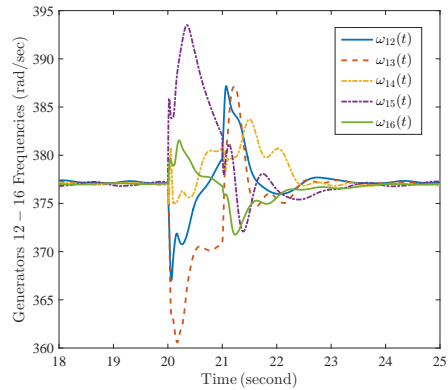}
\caption{Estimated frequencies of the Generators 12--16 before, during, and after a CA. The figure illustrates the impact that the DRMOP has on the estimation of frequencies of generators in a power network. }	\label{fig:frequencies}
	\vspace{-0.2cm}
\end{figure}

Since significant frequency deviations are remediated by corrective actions through feedback, e.g., re-dispatch or AGC-like control strategies, misleading values for a generator's frequency deviations can lead to erroneous control actions. Fig.~\ref{fig:frequencies} shows the frequencies of Generators 12--16, before, during, and after a CA, depicting the impact that the DRMOP has on the estimation of frequency deviations given UIs. As illustrated in Fig.~\ref{fig:frequencies}, without near-immediate identification and detection of attacked measurements, wrong corrective actions might be taken. This can steer the system to instability, while the system is in reality not experiencing high frequency deviations; rather, it is experiencing a CA against measurements.

From Fig.~\ref{fig:frequencies}, and prior to a CA, the frequencies are $60$ Hz ($377$ rad/sec). During a CA, the estimated frequencies of the five generators indicate serious system instability, which could be wrongly mitigated by feedback control actions. As discussed in the previous section, the DRMOP succeeds in identifying the wrong measurements, and the frequencies are stable again---indicating a normal grid-condition. Note that the fluctuations are due to the unknown input vector $\m w(t)$. 

\section{Closing Remarks \& Future Work}~\label{sec:concl}
PMUs can be utilized for ameliorated monitoring and control of the smart grid.
The humongous size of data generated by PMUs---communicated to the operators in control centers---can be used for a more accurate depiction of how power system is behaving, and occasionally \textit{misbehaving}.
\subsection{Paper Contributions \& Future Work}
For ameliorated risk mitigation due to CAs in power networks, the focus of the research in this paper is on the unique, seamless integration of three intertwined components:
\begin{itemize}
	\item First, we utilize a robust dynamic state estimator for perturbed grid dynamics. Then, estimates of the system's UIs and possible CAs are obtained. These estimates are then utilized for attack detection and isolation.
	\item Second, the state- and UI estimation components are utilized in an optimization scheme to determine the\textit{ most faulty measurements} with an attack detection filter. 
	\item Finally, a risk mitigation strategy is employed to guarantee a minimal threat-level, ensuring the observability of the power system through available \textit{safe} measurements. 
\end{itemize}   Our future work in this area will focus on three main tasks:
\begin{itemize}
	\item Deriving closed-form solutions for \textit{simultaneous} UIs and CAs estimates for a nonlinear power system model;
	\item Extending the dynamic risk mitigation problem by incorporating conventional devices and accounting for the probabilistic threat-levels in PMU networks as in~\cite{Amir2015}.
	\item Developing a computationally superior method for faster dynamic state estimation for a power system, leveraging the inherent sparsity of the power network.
\end{itemize}

\subsection{What this Paper does not Address; Potential Solutions}
While the methods investigated in this paper address the interplay between DSE, unknown inputs reconstruction, and cyber-attacks detection and mitigation, 
there are two main challenges that are not addressed.

The first challenge is the DSE for a nonlinear representation of the power system. In this paper, we use a linearized representation for the power system as:
		\begin{itemize}
			\item[(a)] Real-time state and UI estimates can be obtained in real-time, whereas for a highly nonlinear power system, these estimates require stricter assumptions and a stronger computational capabilities that guarantee  the convergence of these estimates to their corresponding actual values;
			\item[(b)] Under dynamic UIs and CAs, the problem of estimating states for 
			higher-order nonlinear power systems has never been addressed before in the literature. 
			The majority of routines are based on Kalman filter derivatives, having major difficulties under unknown inputs given nonlinear systems. For the risk mitigation for nonlinear systems, efficient computational solutions will have to be developed to ensure near real-time guarantees of the system's observability---this is a research question related to our work in this paper, but is beyond this paper's objectives.
			\end{itemize}
The second challenge is the incorporation of dynamic loads. We have assumed that the unknown inputs incorporate unknown quantities such as model uncertainties, linearization errors, and disturbances. We also illustrated that the derived estimator successfully estimates these unknown quantities without prior knowledge on their distribution. However, variable loads may not be immediately incorporated in the linearized dynamic model, and if heavy loads are experienced during a cyber-attack, the linear system model might become invalid.

To address these challenges, polytopic representations of power systems can be derived that capture various linearized representations corresponding to various loading conditions (for different time-periods), whereby the SMO can be designed given these conditions. Consequently, a unique observer gain can be computed that stabilizes the estimation error norm for all time-periods, considering variable loads and dynamic unknown inputs. In particular, a common $\m L_q$ can be solved for different pairs of $(\mA,\mC_p)$ matrices for different time-periods. This idea is analogous to designing a common controller gain for switched dynamical systems.

\bibliographystyle{ieeetran}
\bibliography{bibfile}
\begin{IEEEbiography}[{\includegraphics[width=1in,height=1.25in,clip,keepaspectratio]{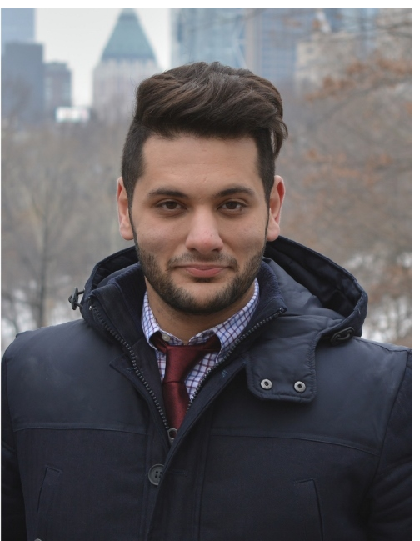}}]{Ahmad F. Taha}
(S'07--M'15) received the B.E. and Ph.D. degrees in Electrical and Computer Engineering from the American University of Beirut, Lebanon in 2011 and Purdue University, West Lafayette, Indiana in 2015. 

In Summer 2010, Summer 2014, and Spring 2015 he was a visiting scholar at MIT, University of Toronto, and Argonne National Laboratory. Currently he is an assistant professor with the Department of Electrical and Computer Engineering at The University of Texas, San Antonio. Dr. Taha is interested in understanding how complex cyber-physical systems operate, behave, and \textit{misbehave}. His research focus includes optimization and control of power system, observer design and dynamic state estimation, and cyber-security.
\end{IEEEbiography}

\begin{IEEEbiography} [{\includegraphics[width=1in,height=1.25in,clip,keepaspectratio]{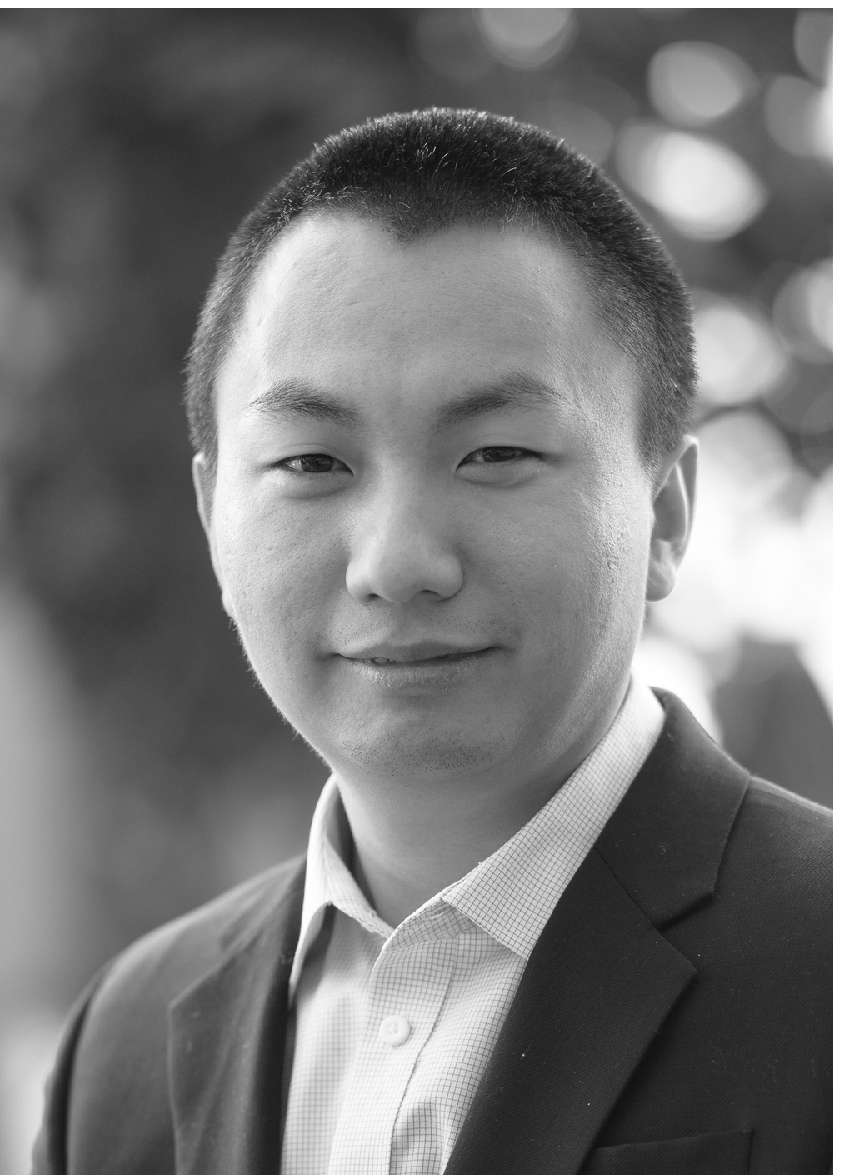}\vfill}]
	{Junjian Qi} (S'12--M'13)
	received the B.E. degree from Shandong University, Jinan, China, in 2008 and the Ph.D. degree Tsinghua University, Beijing, China, in 2013, both in electrical engineering.
	
	In February--August 2012 he was a Visiting Scholar at Iowa State University, Ames, IA, USA. During September 2013--January 2015 he was 
	a Research Associate at Department of Electrical Engineering and Computer Science, University of Tennessee, Knoxville, TN, USA. 
	Currently he is a Postdoctoral Appointee at the Energy Systems Division, Argonne National Laboratory, Argonne, IL, USA. 
	His research interests include cascading blackouts, power system dynamics, state estimation, synchrophasors, and cybersecurity.
\end{IEEEbiography}

%
%

\begin{IEEEbiography} [{\includegraphics[width=1in,height=1.25in,clip,keepaspectratio]{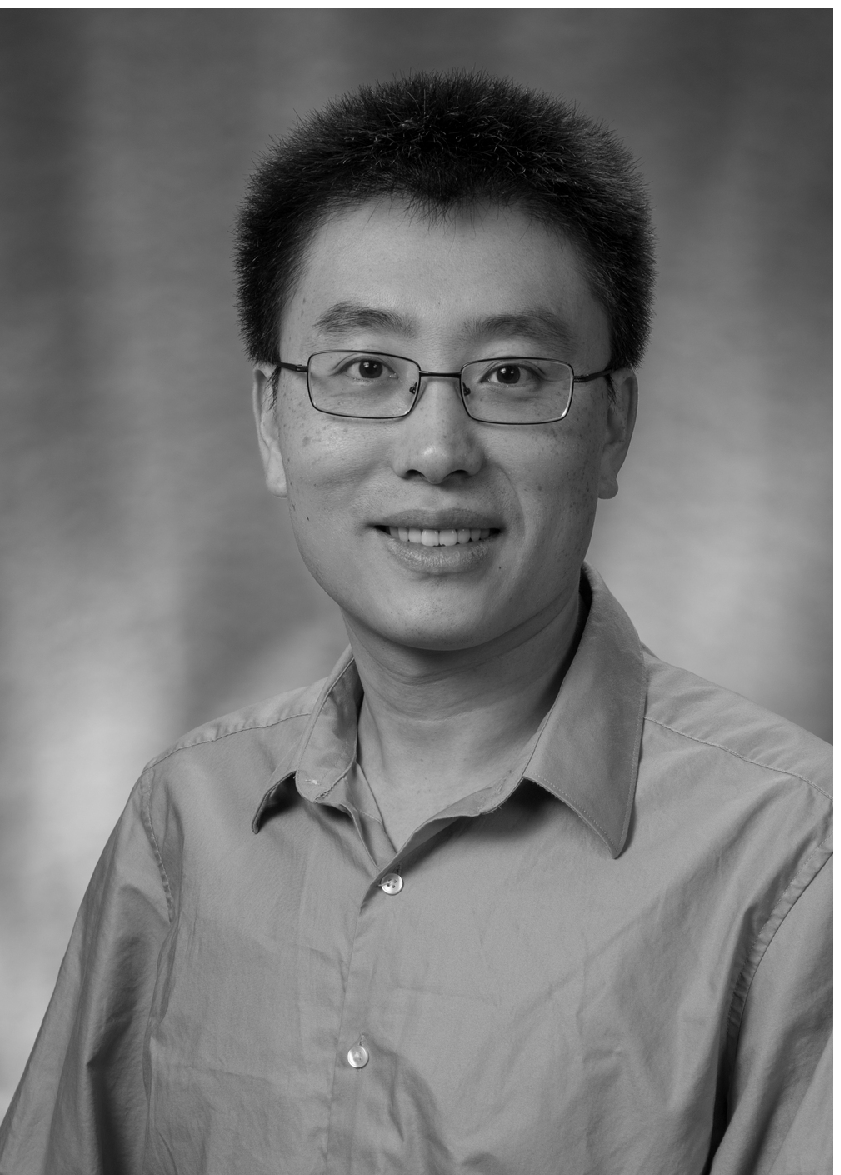}\vfill}]
	{Jianhui Wang} (S'07--SM'12) 
	received the Ph.D. degree in electrical engineering from Illinois Institute of Technology, Chicago, IL, USA, in 2007. 
	
	Presently, he is the Section Lead for Advanced Power Grid Modeling at the Energy Systems Division at Argonne National Laboratory, Argonne, IL, USA.
	Dr. Wang is the secretary of the IEEE Power \& Energy Society (PES) Power System Operations Committee. 
	
	He is an Associate Editor of Journal of Energy Engineering and an editorial board member of Applied Energy. He is also an affiliate professor at Auburn University and an adjunct professor at University of Notre Dame. He has held visiting positions in Europe, Australia, and Hong Kong including a VELUX Visiting Professorship at the Technical University of Denmark (DTU). Dr. Wang is the Editor-in-Chief of the IEEE Transactions on Smart Grid and an IEEE PES Distinguished Lecturer. He is also the recipient of the IEEE PES Power System Operation Committee Prize Paper Award in 2015.
\end{IEEEbiography}

\begin{IEEEbiography}[{\includegraphics[width=1in,height=1.25in,clip,keepaspectratio]{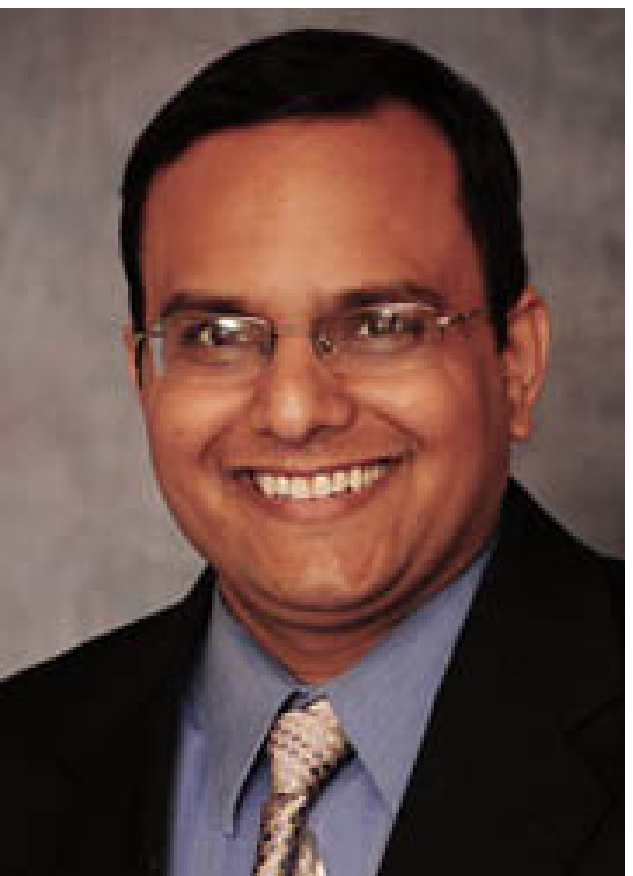}}]{Jitesh H. Panchal}
 (S'00--M'05) received his BTech (2000) from Indian Institute of Technology (IIT) Guwahati, and MS (2003) and PhD (2005) in Mechanical Engineering from Georgia Institute of Technology.  
 
 He is an Associate Professor in the School of Mechanical Engineering at Purdue University. Dr. Panchal's research interests are in the science of systems engineering with focus on three areas: (i) democratization of design and manufacturing, (ii) decision making in decentralized socio-technical systems, and (iii) integrated products and materials design. 
 
Dr. Panchal is a co-author of the book titled Integrated Design of Multiscale, Multifunctional Materials and Products. He is a recipient of CAREER award from the National Science Foundation (NSF), Young Engineer Award and two best paper awards from ASME CIE division, and a university silver medal from IIT Guwahati.
\end{IEEEbiography}

\end{document}